\documentclass{aa}
\usepackage{graphicx}
\usepackage{epsfig}
\usepackage{txfonts}
\newcommand{\mincir}{\raise
-3.truept\hbox{\rlap{\hbox{$\sim$}}\raise4.truept\hbox{$<$}\ }}
\newcommand{\magcir}{\raise
-3.truept\hbox{\rlap{\hbox{$\sim$}}\raise4.truept\hbox{$>$}\ }}
\begin{document}
   \title{Classification and environmental properties of X-ray
     selected point-like sources in the XMM-LSS field}


   \author{O. Melnyk \inst{1,2}, M. Plionis\inst{3,4}, A. Elyiv\inst{1,5}, M. Salvato\inst{6,7},
      L. Chiappetti\inst{8}, N. Clerc\inst{6}, P. Gandhi\inst{9},
      M. Pierre\inst{10}, T. Sadibekova\inst{10},
      A. Pospieszalska-Surdej\inst{1} and J. Surdej\inst{1}}

   \offprints{O. Melnyk}

   \institute{Institut d'Astrophysique et de G\'eophysique,
     Universit\'e de Li\`ege, 4000 Li\`ege, Belgium\\
  \email{melnykol@gmail.com}
	 \and
     Astronomical Observatory, Taras Shevchenko National University of Kyiv, 3
     Observatorna St., 04053 Kyiv, Ukraine\	  \and
    Physics Dept., Aristotle Univ. of Thessaloniki, Thessaloniki 54124,
    Greece \and
  Instituto Nacional de Astr\'ofisica, \'Optica y Electr\'onica,
72000 Puebla, Mexico \and
Main Astronomical Observatory, Academy of Sciences of Ukraine, 27
Akademika Zabolotnoho St., 03680 Kyiv, Ukraine \and
Max-Planck-Institute for Extraterrestrial Physics, Giessenbachstrasse 1 , Garching, 85748 Germany \and
Excellence Cluster, Boltzmannstrass 2, 85748 Garching, Germany \and
INAF-IASF Milano, via Bassini 15, I-20133 Milano, Italy \and
 Institute of Space and Astronautical Science (ISAS), Japan
 Aerospace Exploration Agency, 3-1-1 Yoshinodai, Chuo-ku, Sagamihara,
 Kanagawa 252-5210, Japan  \and
DSM/Irfu/SAp, CEA/Saclay, F-91191 Gif-sur-Yvette Cedex, France \\
} 


\date{Received Xxxxx xx, 2010; accepted Xxxx xx, 2010}


 \abstract
   {The XMM-Large Scale Structure survey, covering an area of 11.1 sq. deg., 
     contains more than 6000 X-ray point-like sources detected with
     XMM-Newton down to a flux of $3\times10^{-15}$ erg s$^{-1}$
     cm$^{-2}$ in the [0.5-2] keV band, the vast majority of which
     have optical (CFHTLS), infrared (SWIRE IRAC and MIPS), near-infrared
     (UKIDSS) and/or ultraviolet (GALEX) counterparts.}
   {We wish to investigate the environmental properties of
     the different types of the XMM-LSS X-ray sources, defining their
     environment using the $i'$-band
     CFHTLS W1 catalog of optical galaxies down to a magnitude limit of 23.5 mag.}
  {We have classified 4435 X-ray selected sources on the basis of their spectra, SEDs and
X-ray luminosity and estimated their photometric redshifts, having 4-11 band
  photometry, with an accuracy $\sigma_{\triangle z/(1+z_{sp})}$=0.076
  and 22.6\% outliers for $i'<$26 mag. We estimated
  the local overdensities of 777 X-ray sources which have spectro-z or
  photo-z calculated using more than 7 bands (accuracy
  $\sigma_{\triangle z/(1+z_{sp})}$=0.061
  with 13.8\% outliers) within the volume-limited
  region defined by $0.1\leq z \leq0.85$ and -23.5$<M_{i'}<$-20. }
  {Although X-ray sources may be found in
     variety of environments, a large fraction ($\magcir$55-60\%), as
     verified by comparing with the random expectations, reside in
     overdense regions. 
     The galaxy overdensities within which X-ray sources reside show 
     a positive recent redshift evolution (at least for the range
     studied; $z\mincir0.85$). 
     We also find that X-ray selected galaxies, with respect to AGN,
     inhabit significantly higher galaxy overdensities, although their
     spatial extent appear to be smaller than that of AGN. 
     Hard AGN ($HR\geq-0.2$) are located in more overdense regions
     with respect to the Soft AGN ($HR<-0.2$), a fact clearly seen in
     both redshift ranges, although it appears to be stronger in the
     higher redshift range ($0.55<z<0.85$).
     Furthermore, the galaxy overdensities (with $\delta\magcir 1.5$) within which Soft
     AGN are embedded appear to evolve more rapidly with respect
     to the corresponding overdensities around Hard AGN.}
     {}


   \maketitle
%

\section{Introduction}

Active galactic nuclei (AGN) are among the most powerful energy
emitters in the Universe and trace the locations of active supermassive black
holes (SMBHs) on the cosmic web. Understanding the nature and evolution of
SMBHs as a function of cosmic time and environment
constitutes an important goal of modern high-energy astrophysics. In particular
the coincidence between the star-formation peak of galaxies at redshifts $z\sim$2-3 and
the corresponding formation peak of high-luminosity AGN/QSOs, appears
to link in an intricate way the cosmic histories of galaxies and black
holes, providing an important input in understanding the
formation and evolution of cosmic structures in the Universe (Warren
et al. 1994; Schawinski et al. 2009; Fanidakis et al. 2012).

It is known that environmental effects (cf. interactions, minor or
major galaxy merging) may affect the physical properties of galaxies
such as their morphological type, color, star-formation rate, nuclear 
activity, etc. (Dressler 1980, Blanton et al. 2003a, Kauffmann et
al. 2004, van der Wel  2008, etc.).  The connection between galaxy and
black hole formation and evolution indicates that there must be a
dependence of the physical properties of AGN and of their triggering
mechanism on environment, which has to be clearly established.	
Important questions are: (a) what is the main determining factor of
the galaxy/AGN
properties: intrinsic evolution or environmental influence? and (b) how
do the host galaxy physical properties and the local environment affect
the black hole fuelling mechanisms? To answer these questions, many
authors have been studying the properties of AGN environment (density,
colors, morphology, etc. of the nearest neighbours) and compared these
properties with those of normal galaxies.

For example, Kauffmann et al. (2004) conclude that for a fixed stellar
mass of galaxies, both star formation and nuclear activity strongly
depend on the environment up to $\sim$1 Mpc. Waskett et al. (2005) did not
find any differences between the environmental properties of AGN and
of a control sample of galaxies in the 0.4$<z<$0.6 redshift range.
A qualitatively similar result was obtained by Li et al. (2006) analyzing the
clustering properties of narrow-line AGN. Gilmour et al. (2007) considered the environment of X-ray
selected AGN in the supercluster A901/2 (at z=0.17) and found that
they prefer dense environments, avoiding however the most overdense and underdense regions.
Similarly, Lietzen et al. (2009) using SDSS data
found that QSOs avoid the densest regions and prefer to reside in the
outskirts of superclusters. Constantin et al. (2008) find that local
($z\mincir 0.09$) AGN are more common in voids than in walls
for a same range of masses and accretion rates. The recent analysis 
of Lietzen et al. (2011) shows that radio-quiet QSOs and Seyfert galaxies prefer
low-density regions contrary to radio galaxies which prefer
more dense environments. 

Silverman et al. (2009), on the basis of X-ray selected AGN in the
COSMOS field, reached the conclusion that AGN prefer to reside in
environments which are similar to those of massive galaxies with
substantial levels of star formation. The AGN with low stellar mass
hosts are located over a wide range of environments but AGN with high
stellar mass hosts prefer low density regions. These results are also
in agreement with Montero-Dorta et al. (2009) who found that Seyferts
and X-ray selected AGN at z$\sim$1 almost do not show environmental
dependencies. Moreover low-redshift LINERs and Seyfert galaxies 
appear to inhabit low density environments contrary to high-redshift
LINERs (z$\sim$1) which favour higher density environments.
Contrary to this, Georgakakis et al. (2007; 2008) 
found that the X-ray population at $z\sim$1 avoids underdense
regions and prefers to reside in groups.
Falder et al. (2010) showed that $z\sim1$ AGN have an excess of galaxy
density within a radius of 200-300 kpc and that the local galaxy density
increases with the radio AGN luminosity, but not with the black hole mass.
Bradshaw et al. (2011) showed that X-ray and
radio-loud AGN, with 1.0$<z<$1.5, are located in significantly
overdense regions, with the former being found in the cluster outskirts.
Furthermore, recent clustering studies of X-ray selected AGN 
  have shown that they reside in group-size dark matter halos with masses
  $M_h\sim 10^{13} h^{-1} M_\odot$, significantly more massive than those
  inhabited by optical QSOs (see Miyaji et al. 2011; Allevato et
  al. 2011, Koutoulidis et al. 2012, and references therein).

According to the unified scheme of Antonucci (1993), AGN of different
types, like Seyferts of type I and II as well as broad and narrow line
QSOs (unobscured and obscured ones), should inhabit similar environments
since these objects only differ due to differences in orientation of their torus 
with respect to the line-of-sight. However the standard
orientation-based AGN unification scheme does not consider any
evolution with redshift of the properties of obscured vs. unobscured
AGN. Is there some observational evidence
establishing a similar environment for the different types of AGN? Statistical studies 
lead to contradictory results.

Koulouridis et al. (2006) found in the very local Universe that the
fraction of Seyfert 2
galaxies with close neighbours is significantly larger than
that of their control sample and of Seyfert 1 galaxies, while their
large scale environment does not show any difference with respect to
their control samples (see also Sorrentino et al. 2006 for similar
results). At the same time, Strand et al. (2008) have shown that higher
luminosity AGN inhabit more overdense environments compared to lower
luminosity AGN out to 2 Mpc. The authors also found that in the
redshift range 0.3$<z<$0.6, type II and type I QSOs present similarly overdense
environments, while the environment of dimmer type I quasars appears
to be less overdense than that of type II quasars.

X-ray selected sources which mainly consist of AGN offer information
about the nature and properties of super massive black holes over a wide redshift range
up to z$\sim$4. Therefore X-ray surveys combined with a multiwavelength
analysis of their host galaxies provide an effective tool for the
environmental study of different types of AGN. In our work we
consider the environmental vs. intrinsic properties of X-ray
selected, but with optically detected counterparts, point-like sources
from the 11.1 sq. deg XMM-Large Scale Structure (LSS) medium-deep extragalactic survey.
Earlier versions of our survey, over a smaller solid angle, and a variety of
analyses can be found in Chiappetti et al. (2005), Gandhi et al. (2006), Pierre et al. (2007),
Tajer et al. (2007), Polletta et al. (2007), Garcet et al. (2007),
Tasse et al. (2008a,b) and Nakos et al. (2009).

In particular, regarding previous environmental studies based on the original
XMM-LSS field, Tasse et al. (2008b) considered the environment of 110 radio-loud AGN finding
that high and low stellar mass systems are located in different
environments. 
The authors concluded that the AGN triggering mechanism of high mass systems
could be produced by cooling of the hot gas and for the low mass systems it can be
explained by the cold gas accretion due to a merging. In addition, Tasse et
al. (2011) found that X-ray selected type 2 AGN show very similar
individual and environmental properties as low mass radio-loud AGN.

The main aim of our current work is to consider the environmental properties
of different types of X-ray selected XMM-LSS sources using the local
density of optical galaxies based on the  CFHTLS\footnote{Canada France Hawaii
  Telescope Legacy Survey: http://www.cfht.hawaii.edu/Science/CFHLS/}.
In our analysis we use the newest version of the XMM-LSS
multiwavelength catalog (Chiappetti et al. 2012) which contains 
6342 X-ray sources\footnote{In the present paper we only considered 
  objects from the "good" fields i.e. with the condition
  "badfield=0".} over 11.1 sq. deg.
 The clustering
properties of the point-like sources of this catalog were analyzed by Elyiv
et al. (2012).

In section 2 we present the XMM-LSS source sample. In Section 3, we
discuss the classification of the optical counterparts, the photo-$z$
determination technique and the corresponding results. Section 4 presents the
multiwavelength properties of the sources and the 
samples of the different types of X-ray sources. The results of the
environmental analysis are given in Section 5. Discussion and
general conclusions form Section 6. Throughout
this work we use the standard cosmology: $\Omega_{0}$=0.3, $\Omega_{\Lambda}$=0.7 and
$H_{0}$=72 km/s/Mpc.

\section {The sample of X-ray sources and their counterparts}
The XMM-LSS field occupies an area of 11.1 sq. deg. and is located
at high galactic latitudes by [2$^{h}14^{m}<R.A.<2^{h}30^{m}$;
-6$\raisebox{1ex}{\scriptsize o}25^{m}<Dec<-2
\raisebox{1ex}{\scriptsize o}35^{ m}$, J2000.0], see Fig. 1. This
field also contains the Subaru X-ray Deep Survey (SXDS) 1.14 sq. deg. deep field (Ueda et
al. 2008). We will refer below to the XMM-LSS field as the ``Full
Exposure Field''\footnote{The data are available in the Milan DB in
  the 2XLSSd and 2XLSSOPTd tables. See Chiappetti et al. (2012) for
  details.} which includes the SXDS field.

The XMM-LSS full exposure field contains 6342 sources$^2$ detected in
the soft (0.5-2 keV) and/or hard (2-10 keV) bands down to a detection
likelihood of 15. This corresponds to the following flux limits (50\%
detection probability): F$_{(0.5-2 keV)}= 3\times10^{-15}$ erg
s$^{-1}$ cm$^{-2}$ and F$_{(2-10keV)}= 1\times10^{-14}$ erg
s$^{-1}$cm$^{-2}$, over nominal survey pointings.

See further
details about the catalog sensitivity in Elyiv et al. (2012) and
the catalog description paper by Chiappetti et al. (2012).
 In the latter paper the details of the soft and hard band "merging"
 are also given.  From the multiwavelength catalog of the X-ray
point-like sources and their counterparts in optical/CFHTLS,
infrared/SWIRE\footnote{Spitzer Wide-area InfraRed Extragalactic
  legacy  survey: http://swire.ipac.caltech.edu//swire/swire.html}/IRAC\footnote{Infrared
  Array Camera on the Spitzer Space Telescope} and MIPS\footnote{The
  Multiband Imaging Photometer for Spitzer}, near-infrared
UKIDSS\footnote{The UKIRT Infrared Deep Sky Survey:
  http://www.ukidss.org/} and ultraviolet GALEX\footnote{The Galaxy
  Evolution Explorer: http://www.galex.caltech.edu/}, we only selected
those sources which have a counterpart at optical wavelength or more wavelengths. The detailed
description of the matching of X-ray sources with their
multiwavelength counterparts is given in Chiappetti et al. (2012).

 We only took one
counterpart for each X-ray source; that which had the best $probXO$ probability
(see definition below) and rank 0 (single counterpart) or 1
(preferred counterpart but there are more than one), see Chiappetti et
al. (2012) for details.  The total number of X-ray selected sources
with corresponding counterparts is 5142.

Each counterpart was visually inspected using CFHTLS $g'$, $i'$, $r'$
images from the 6$^{th}$ release by two independent inspectors. The
majority of the counterparts have  $g'$ and $r'$-band observations: 5078
sources are visible in $g'$ and 5047 in the $r'$-band,
respectively. According to our rough visual classification the sources
were split into extended sources ($ext$), point-like sources
($ptl$), the latter consisting mostly of QSOs, stars. We reserved one more category
for very faint/invisible objects, photo-defects, 
etc. These objects are referred as misclassified objects $mcl$. The top
and middle panels of Fig. 2 show the distribution of sources as a
function of their apparent $r'$ magnitude according to the two
independent inspectors. The second inspector tends to classify faint
objects and objects with bright nuclei as $ptl$, while the first
inspector classifies most of these objects as extended ones ($ext$).
Similar types were assigned for about 90\% of objects up to $\sim$22
mag by both inspectors (see the bottom panel of Fig. 2).

Therefore we preliminary considered 2441optical counterparts as $ext$ (if at least one
inspector marks the object as $ext$), 2280 as $ptl$ (if both
inspectors classify the object as $ptl$) and rejected from any further
consideration 238 $mcl$ and 183 stars (83 of them are
spectroscopically confirmed and the rest confirmed from their
SEDs, see section 3).
 From both visual classifications it is clearly seen that stars
typically have magnitudes less than $\sim$16 mag and
$mcl$ counterparts are dominating the sources with magnitudes above
$\sim$25 mag. Such a kind of visual inspection was necessary to reject
"bad" counterparts and to have an idea on how to choose the
photometric templates. The final classification of the sources was
made on the basis of spectroscopic information, SEDs and/or X-ray
luminosities (see section 3 and subsection 4.1 for the details).

In the XMM-LSS catalogs (Chiappetti et al. 2005, 2012, Pierre et
al. 2007), each selected multiwavelength counterpart of an X-ray source is
assigned a probability $probXY=1-\exp(-n(<m)r^{2})$, where $r$ is the angular
distance between the X-ray source and its counterpart, and $n(<m)$ is
the density of objects brighter than the magnitude $m$ of the
counterpart. The random probability of association between
an X-ray source and a counterpart $Y$ is given by $probXY$ (Downes et
al. 1986), where $Y=O, I, ..$ for an Optical, Infrared, etc. counterpart.
The median value of $probXO$ for $ext$, $ptl$, stars and $mcl$ objects is
0.018, 0.009, 0.0002 and 0.109, respectively. Tajer et al. (2007) and
Garcet et al. (2007) ranked the probability of counterparts as "good"
($probXY<0.01$), "fair" ($0.01<probXY<0.03$) and "bad" ($probXY>0.03$). So we
see that our $mcl$ objects naturally fall in the latter category.

\begin{figure}
\includegraphics[width=8.5cm,trim=0 0 35 10,clip]{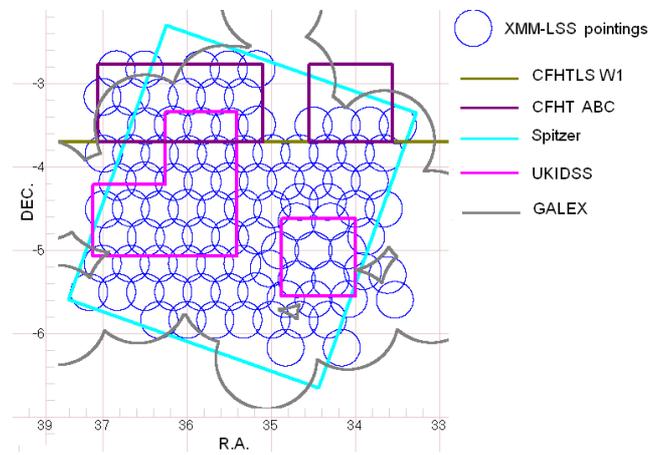}
\caption{XMM-LSS field and multiwavelength coverage. CFHT ABC
  supplementary pointings are not the part of the Legacy Survey.}
	 \label{1}
      \end{figure}

\begin{figure}
\includegraphics[width=8.5cm,trim=15 15 20 20,clip]{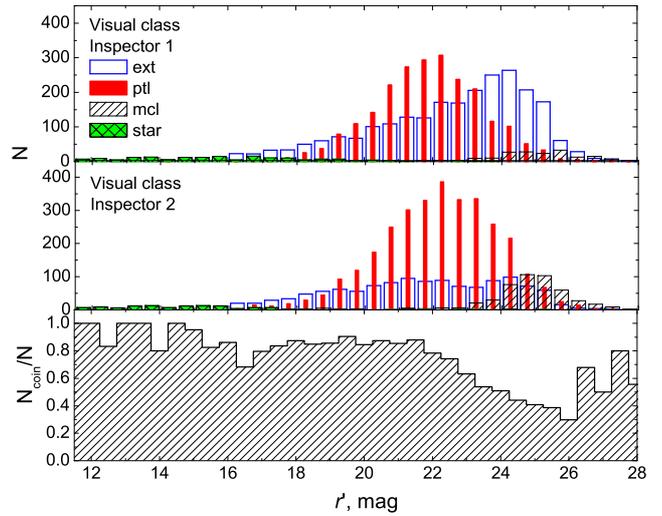}
\caption{Magnitude distribution of 5142 visually classified  objects
  by two independent inspectors (the top and middle panels). The
  fractional agreement between the two independent visual
  classifications is illustrated in the bottom panel.}
\label{2}
\end{figure}

\section{Photo-z determination}
For the photo-z determination we used the
LePhare\footnote{http://www.cfht.hawaii.edu/~arnouts/LEPHARE/lephare.html}
public code (Arnouts et al. 1999, Ilbert et al. 2006).
 First of all we compiled a training sample of objects with known
spectroscopic redshifts.  The spectroscopic redshifts for the XMM-LSS field sources
were taken from the papers which are listed in the description of columns for Table 2. 
We only took into consideration those redshifts for which the angular
separation between the optical counterpart of the X-ray source and the
object observed spectroscopically is less than $1''$. We also
visually verified that the coordinates associated to the spectra
correspond to those of our
optical counterparts. We found 783 spectroscopic redshifts for the
optical counterparts, classified as galaxies and AGN/QSOs. We did not
take into account 51 of the redshifts which have a rank "3", see the
description of columns for Table 2. We have to note that an
inhomogeneity of the spectroscopic data could affect the final
accuracy of the photo-spectro-z relation but at this time we used all available
information.

\begin{figure*}
\begin{tabular}{cc}
\epsfig{file=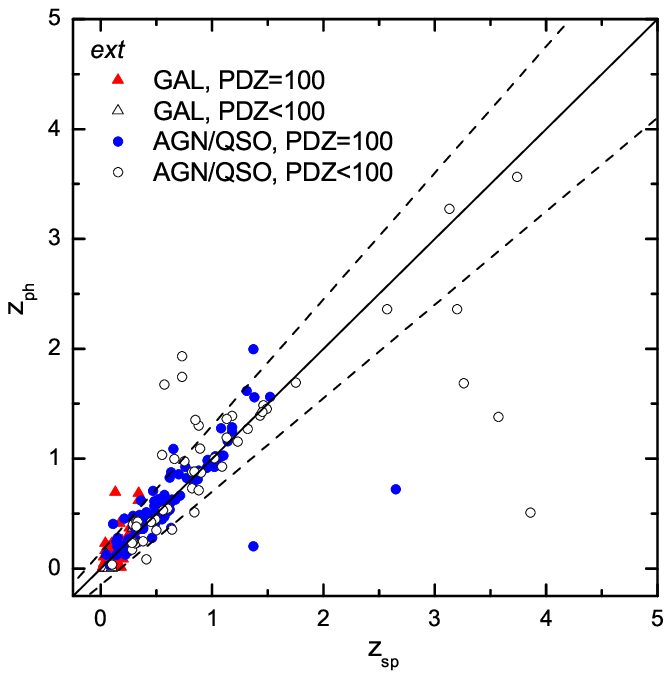,width=8.5cm,trim=15 15 27 25,clip} &
\epsfig{file=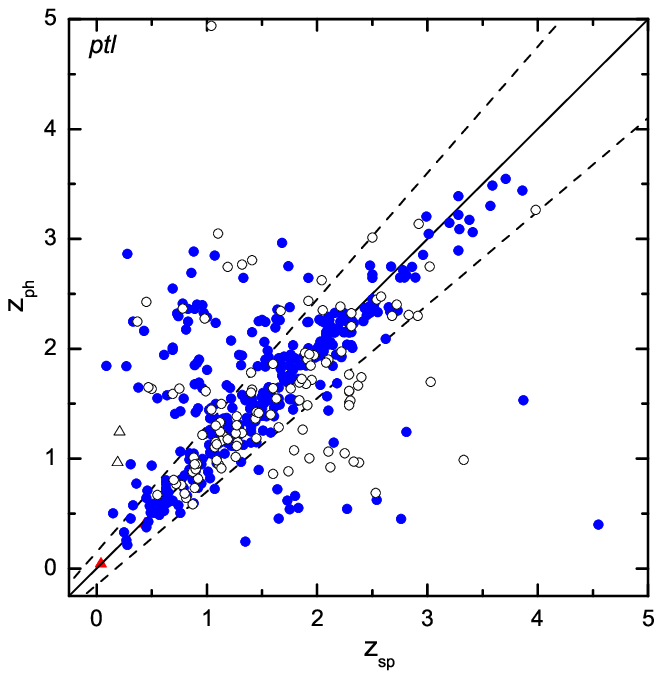,width=8.5cm,trim=15 15 27 25,clip} \\
\end{tabular}
\caption{Photo-$z$ (7+ bands) vs. spectro-$z$ for the 243 visually
  classified extended objects (left panel) and the 479 $ptl$ objects
  (right panel). The solid lines correspond to $z_{ph}=z_{sp}$, the
  dashed lines correspond to $z_{ph}=z_{sp}\pm 0.15\times
  (1+z_{sp})$. PDZ represents the parameter appearing in the
  probability distribution of $z$ given by LePhare. PDF=100 means that
  the solution is unique.}
\label{3}
\end{figure*}

\begin{figure*}
\begin{tabular}{cccc}
\epsfig{file=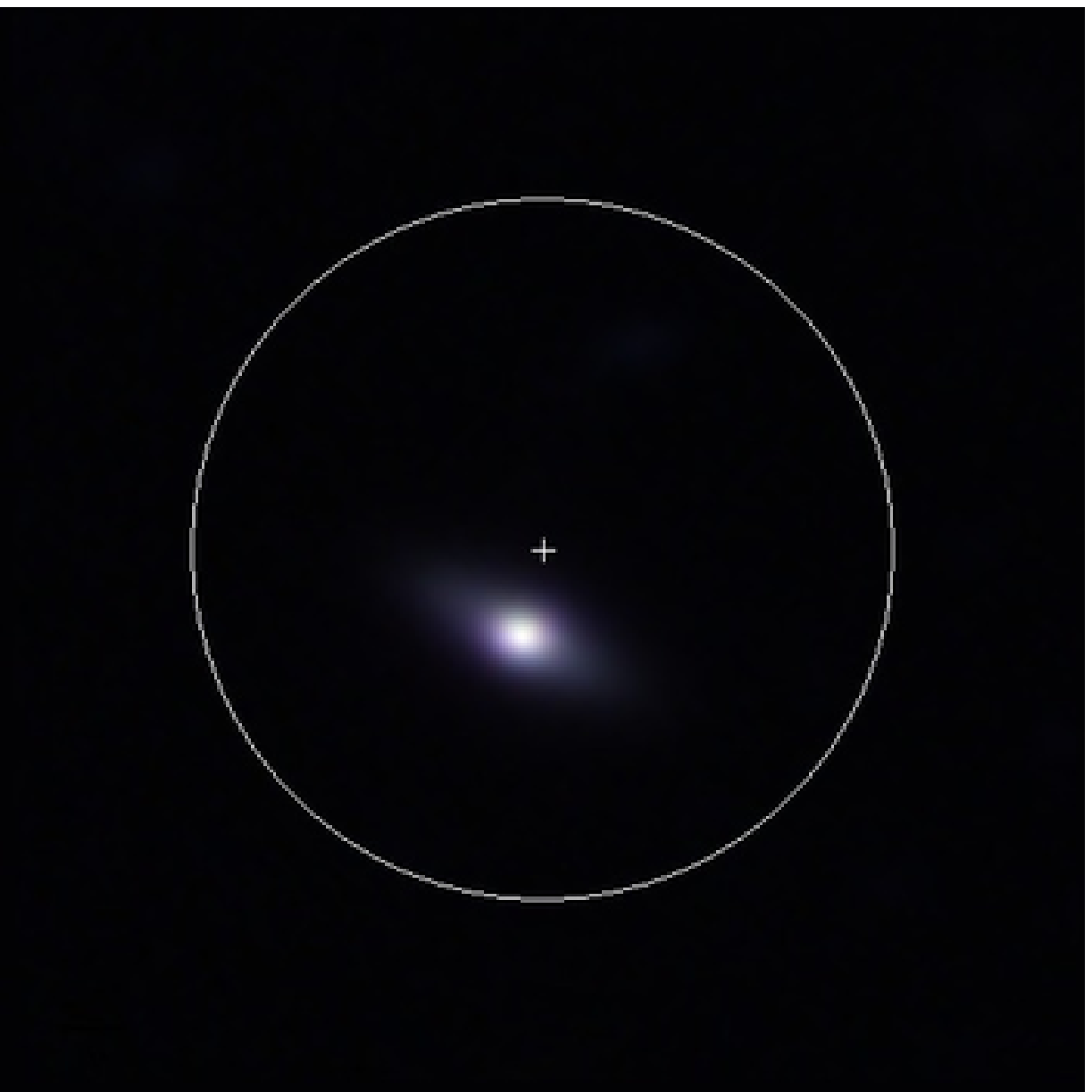,width=4cm} &
\epsfig{file=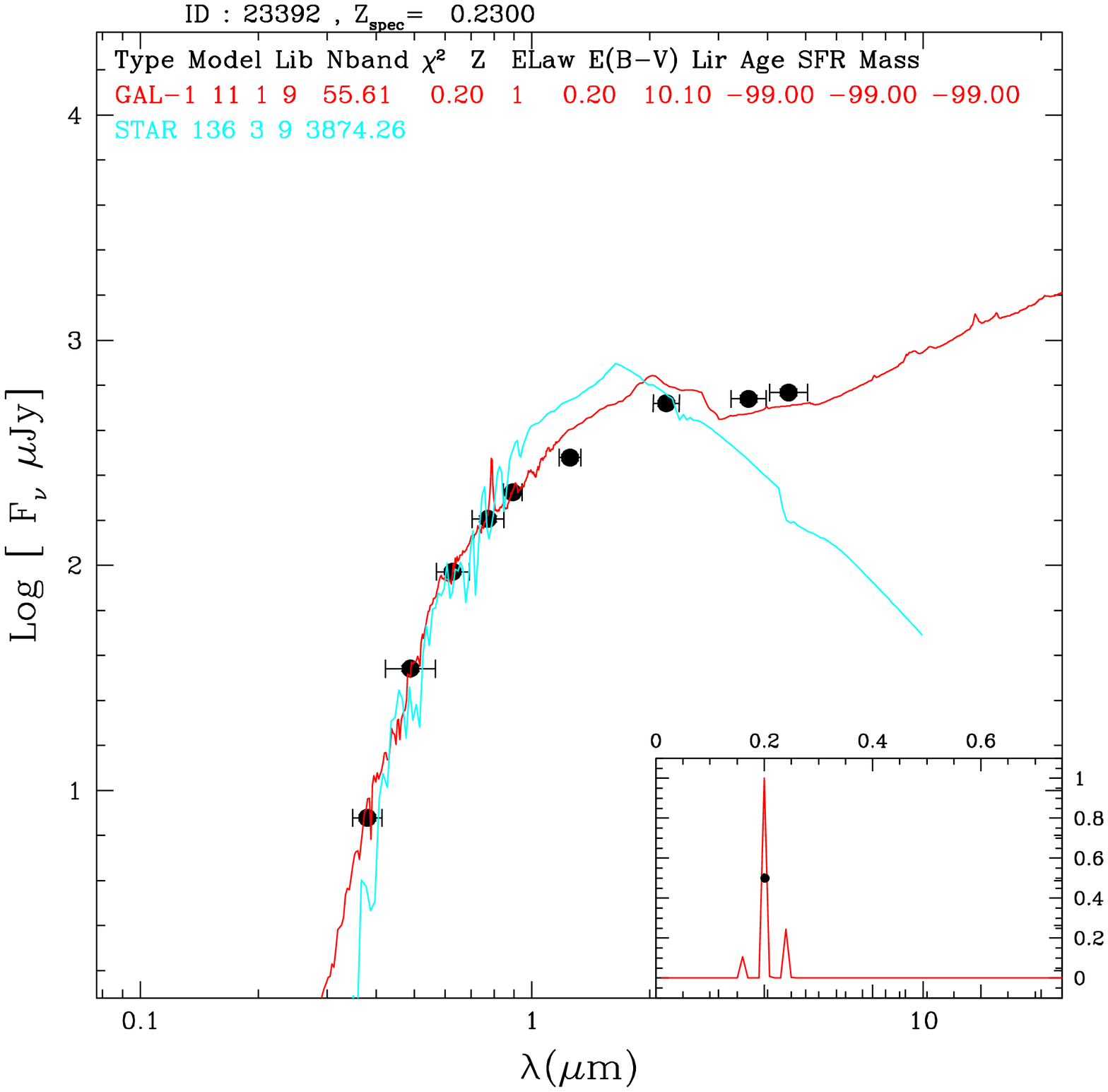,width=4cm} &
\epsfig{file=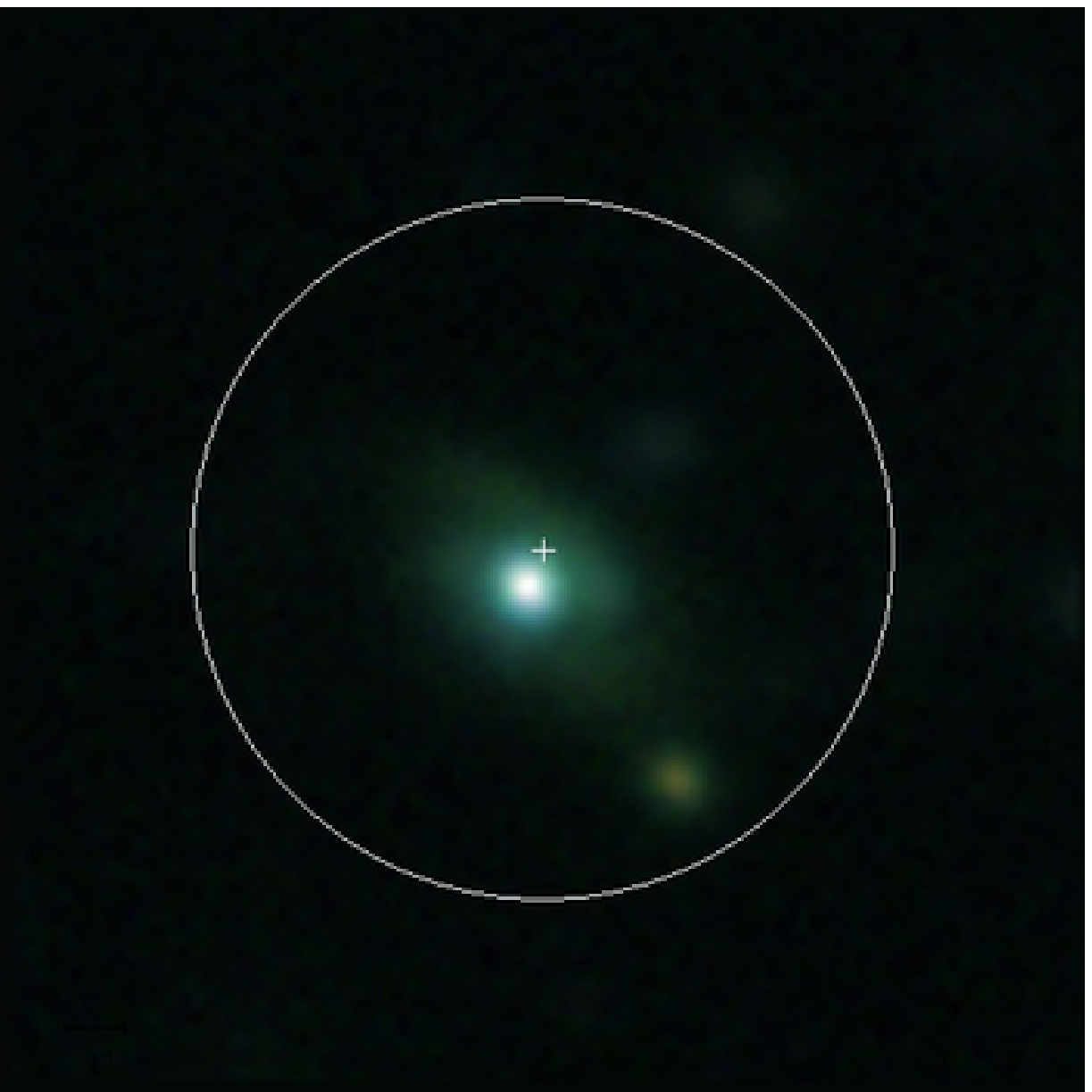,width=4cm} &
\epsfig{file=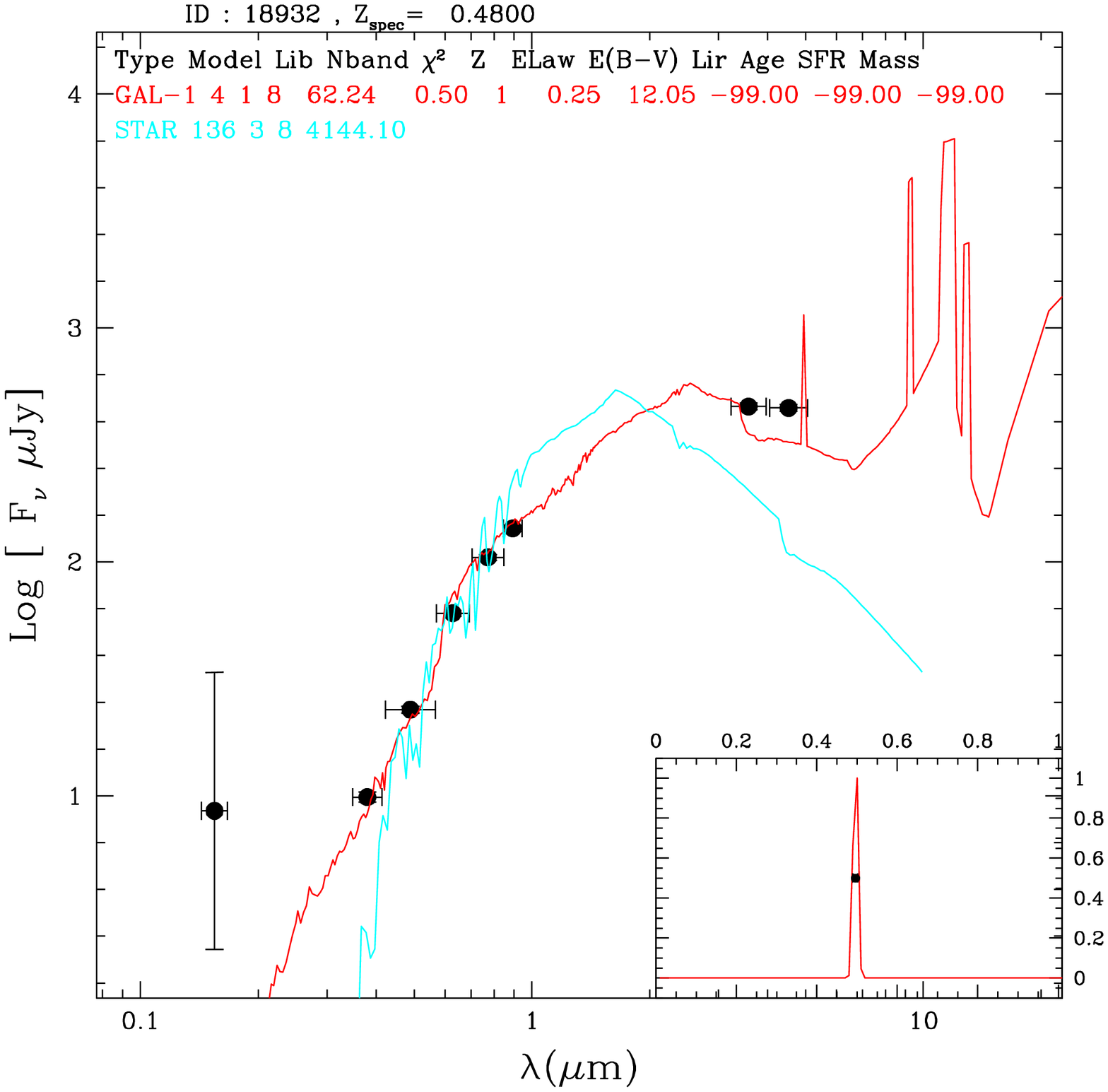,width=4cm} \\
a & b & c & d \\
 \epsfig{file=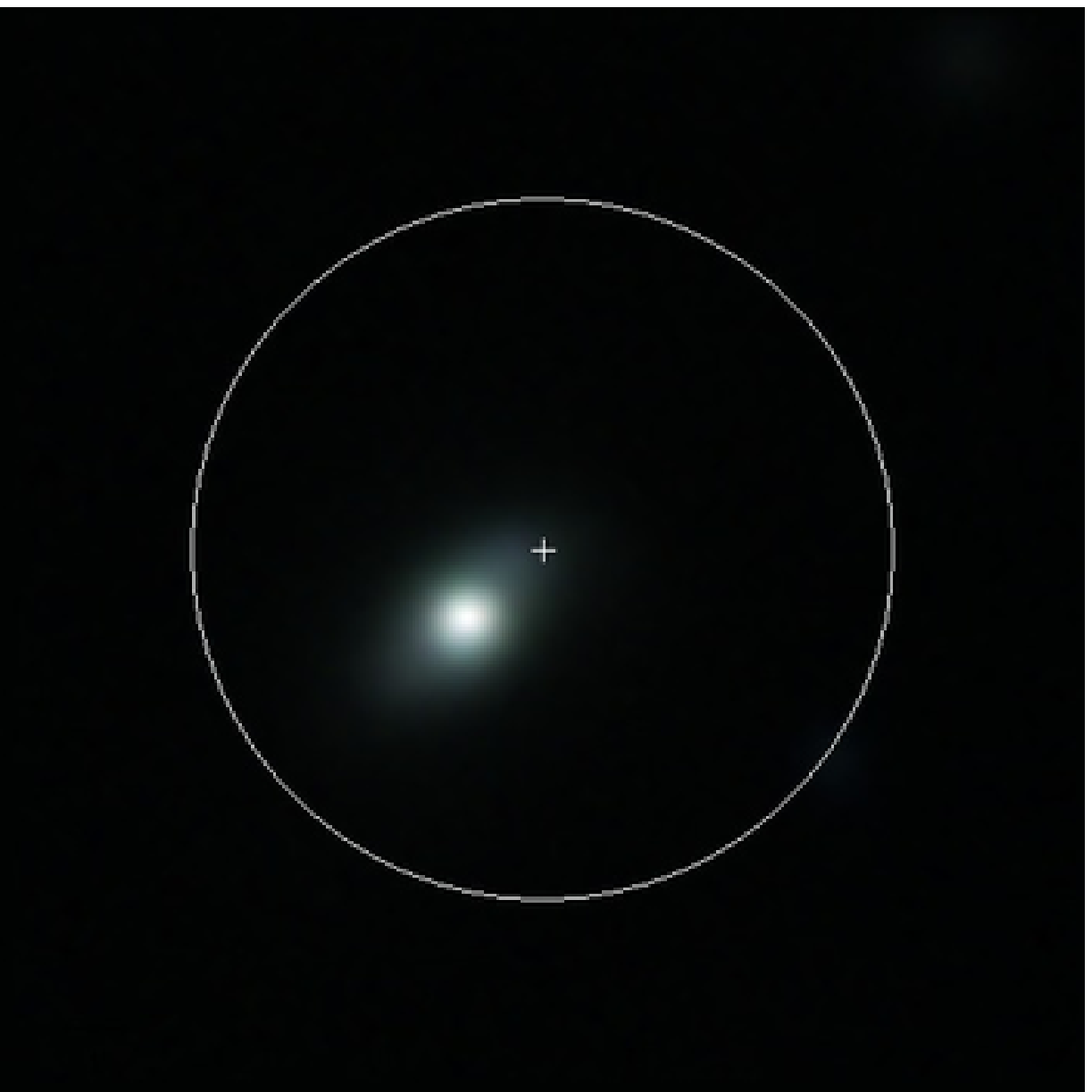,width=4cm} &
 \epsfig{file=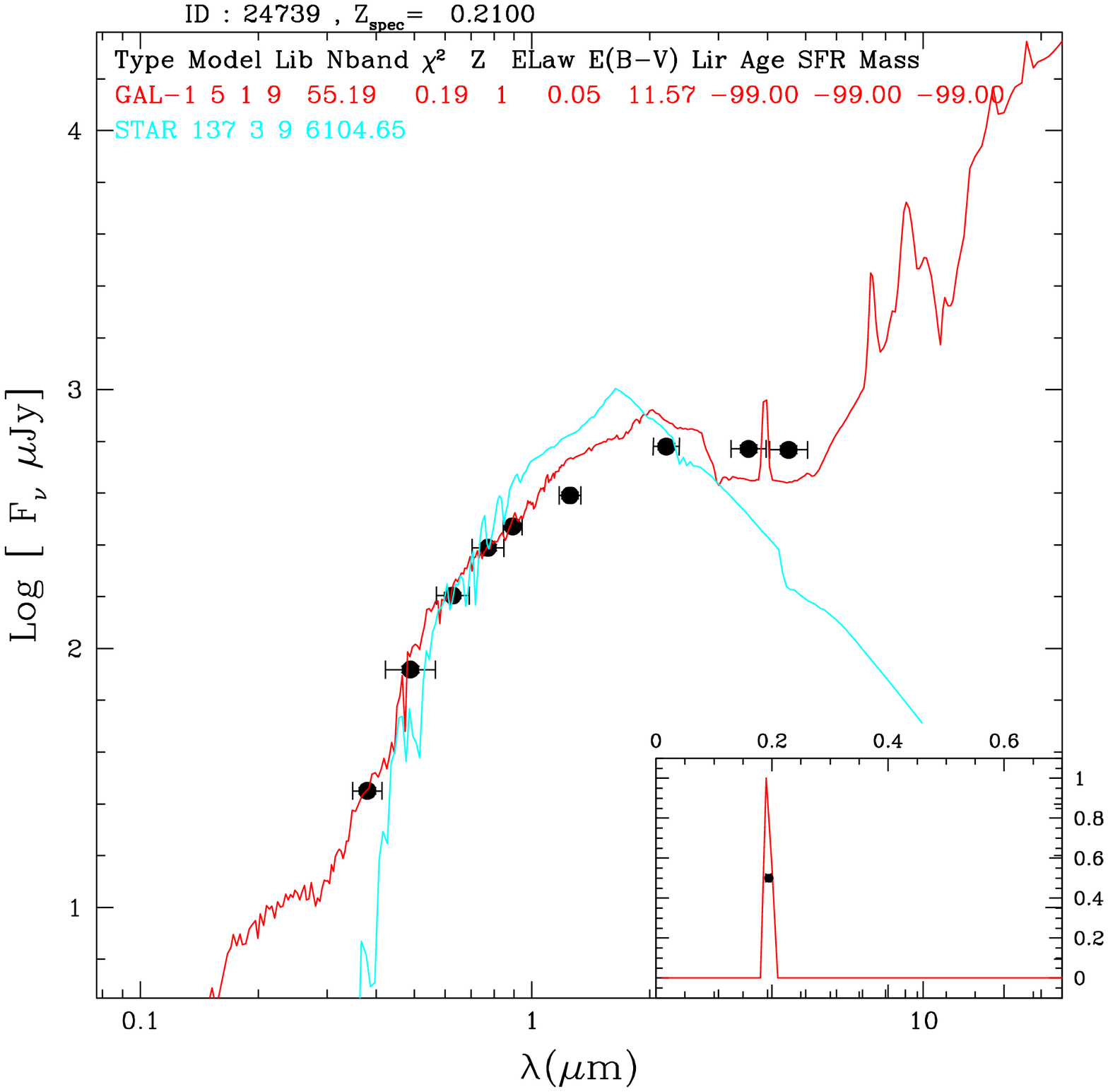,width=4cm} &
 \epsfig{file=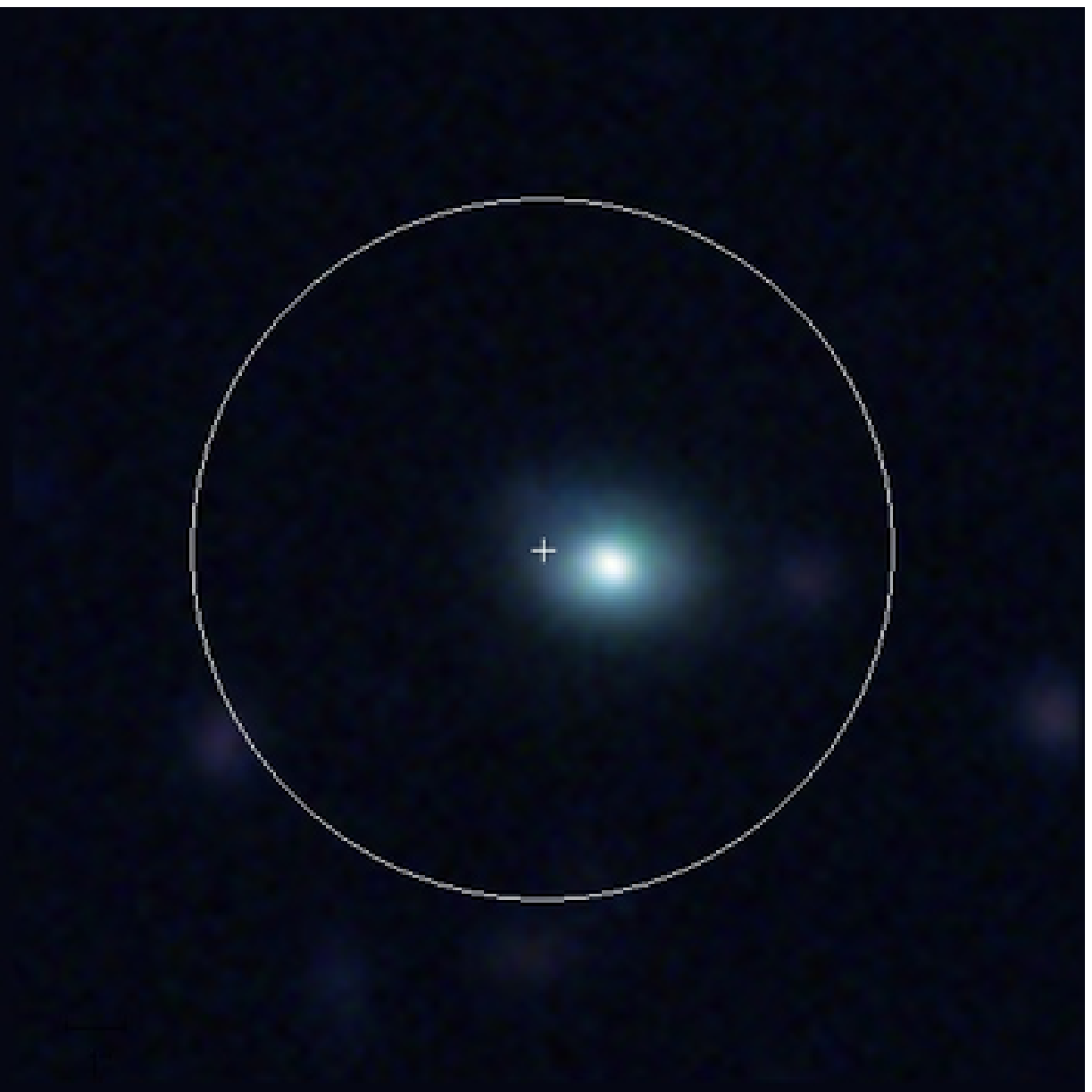,width=4cm} &
 \epsfig{file=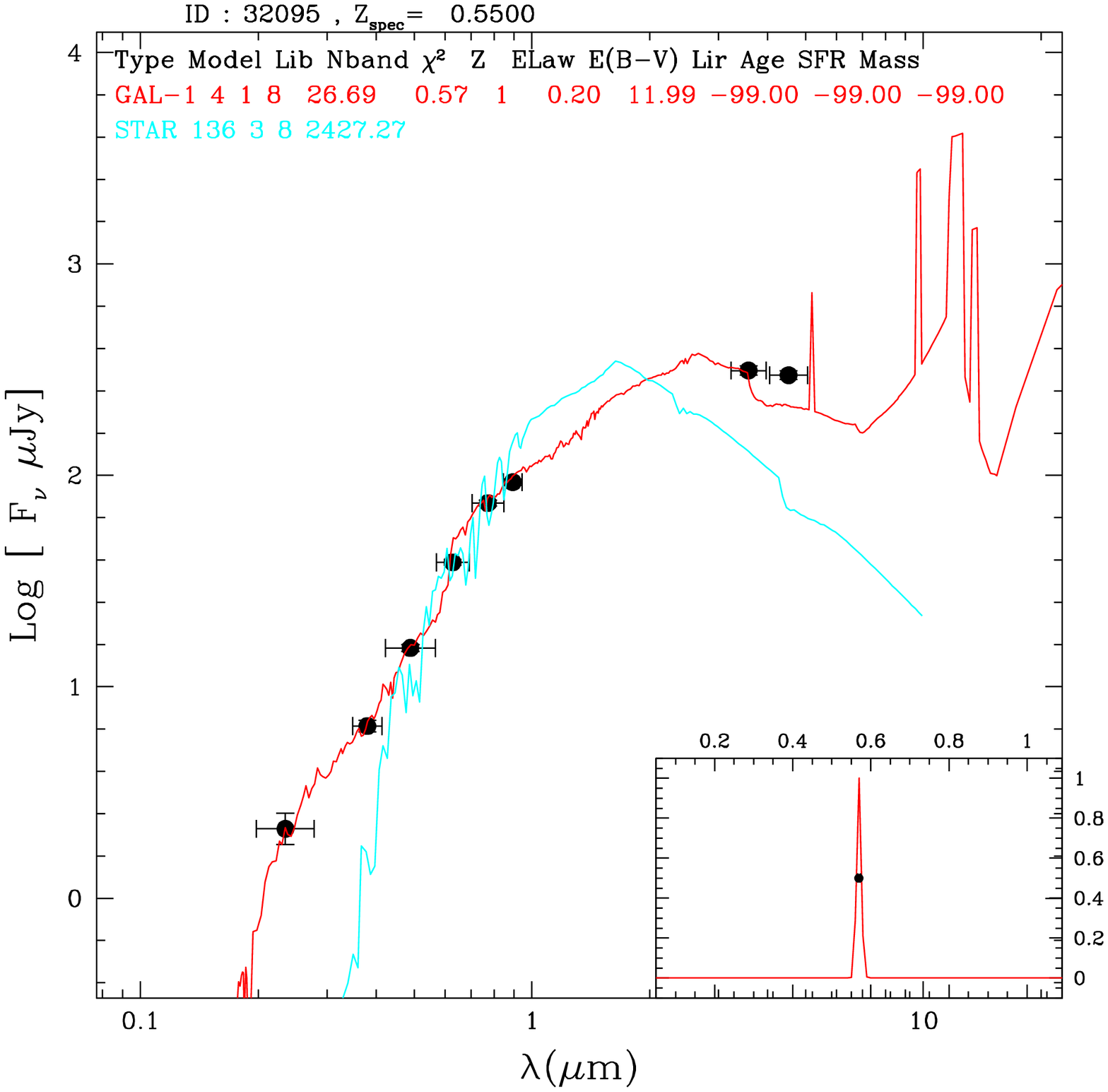,width=4cm} \\
e & f & g & h \\

\end{tabular}
\caption{Examples of optical CFHTLS images and SEDs for selected objects
  from the $ext$ sample fitted with hybrid AGN contaminated templates. The color
  images are composed of the $g'$, $i'$ and $r'$ band images. Below we
  list the spectro-$z$, the source of the spectral classification and
  estimated photo-$z$:  $z_{sp}$=0.23 by Stalin et al. (2010) ,
  $z_{ph}$=0.15 (a,b);
$z_{sp}$=0.478 by Simpson et al. (2012), $z_{ph}$=0.496 (c,d);
$z_{sp}$=0.205 by Garcet et al. (2007), $z_{ph}$=0.195 (e,f);
$z_{sp}$=0.550 by Simpson et al. (2012), $z_{ph}$=0.569  (g,h). The
radius of the white circle is 6$''$.}
\label{4}
\end{figure*}

The photo-$z$ calculation was performed using 11 bands: $u^{*}$, $g'$, $r'$, $i'$, $z'$
(CFHTLS), $J$ and $K$ (UKIDSS), 3.6 and 4.5 $\mu$m (Swire/IRAC)
and far-UV and near-UV bands of GALEX. As recommended  by  the authors
of LePhare, we did not use  bands above $\lambda >  5 \mu$m. Indeed, at those wavelength,
 the templates used for the Spectral Energy
Distribution (SED) fitting are not reliable.  For all the sources of
the sample, we used the total magnitudes
for all bands in the AB photometric system and took into account the
Galactic extinction using $E(B-V)$ values according to the Schlegel
(1998) maps and the Cardelli (1989) extinction law with $R_{v}$=3.1.

As described in Salvato et al (2011), we separately considered two
samples of objects, according to our visual classification: extended, $ext$ (we
assume that this sample mostly consists of
galaxy dominated objects), and point-like, $ptl$ (AGN/QSO dominated
sample).  The photometric redshift calculation was performed using the
Salvato et al. (2009) templates for  our $ptl$ and $ext$ objects with
$F_{(0.5-2keV)}>8\times 10^{-15}$ erg s$^{-1}$ cm$^{-2}$ and Ilbert et
al. (2009) templates for the rest of the fainter X-ray objects
following the flow-chart (Fig.8) by Salvato et al. (2011).  The only
differences are that we did not apply any variability correction and
used the prior -30$<M_{abs}<$-22 for $ptl$. Extinction  laws according
to Calzetti et al. (2000) and Prevot et al. (1984)  were applied to
the Ilbert et al. (2009) normal galaxy templates as free fitting
parameters. To the Salvato et al. (2009) templates we applied only the
Prevot et al. (1984) extinction law. We computed the intrinsic
galactic absorption 
$E(B-V )$  from 0.00 to 0.40 by steps of 0.05. For the redshift
calculation we used a range from 0 to 6
with a step $\Delta z = 0.01$, and at redshifts 6--7 with a step
$\Delta z = 0.05$. We added the emission lines to normal galaxy
templates as in Ilbert et al. (2009).

We defined stars following the condition by Salvato et al. (2011):
1.5$\times \chi^{2}(STAR)<\chi^{2}(AGN/GAL)$, where $\chi^{2}(STAR)$
and $\chi^{2}(AGN/GAL)$ are the reduced $\chi^{2}$ for the best-fit
solutions obtained with stellar and AGN or galaxy libraries. We found
183 stars, 83 from these were spectroscopically confirmed.

We estimated the photometric redshift accuracy
using the measure $\sigma_{\Delta z/(1+z_{sp})}$, which according
to Hoaglin et al. (1983) is given by:
\begin{equation}
\sigma_{\Delta z/(1+z_{sp})}=1.48 \times \frac{{\rm median}|z_{ph}-z_{sp}|}{(1+z_{sp})}
\end{equation}
The outliers are defined as those objects for which
$|(z_{ph}-z_{sp})/(1+z_{sp})|>$0.15.

\begin{table*}
\caption{Photometric redshift accuracy and fraction of the outliers
  for the samples of selected objects which have different numbers of
  photometric bands: 4 and more (4+), etc.}

\tabcolsep9pt
\begin{tabular}{lccc|ccc|ccc} \hline

\hline
 & \multicolumn{3}{c}{	$ext$} &		\multicolumn{3}{c}{$ptl$} & \multicolumn{3}{c}{$all$} \\	
Sample &	N$_{tot}$ (N$_{sp}$) &	 $\sigma_{\Delta
  z/(1+z_{sp})}$   &	$\eta$,\%  &	N$_{tot}$ (N$_{sp}$)  &
$\sigma_{\Delta z/(1+z_{sp})}$ &	$\eta$,\% &	N$_{tot}$
(N$_{sp}$) &	 $\sigma_{\Delta z/(1+z_{sp})}$	 & $\eta$,\% \\
\hline
4+ bands  &	2234(259) & 0.057 &11.2 &	2201 (505) &	0.091 &	27.6 &	4435 (764) &	0.076 &	22.6 \\
7+ bands & 1330 (243) & 0.057 &	10.4 &	1725 (500) &	0.087	 & 26.7	 &3055	 (718)	 &0.074 &	21.8 \\
9+ bands &	415 (130) & 0.057 &	7.7 &	570 (226) & 0.074	 & 22.1 &	985 (356) &	0.063 &	16.9 \\
7+ and PDZ=100 &	743 (195) &	0.052 & 7.3	 & 1072 (373) & 0.074 & 23.9  & 1815  (568)  & 0.065  &	18.1 \\
7+ and $i'<$22  &	819 (209)  &	0.057	 & 8.1  & 1278	  (334)  & 0.076  & 25.0  & 2097  (635)  & 0.071  & 	20.0 \\
\hline
\end{tabular}
\end{table*}

In Table 1 we show the values of the accuracy $\sigma_{\Delta
  z/(1+z_{sp})}$  and of the fraction of the outliers $\eta$  for the
different samples of $ext$, $ptl$ and $all$=$ext$+$ptl$ objects as
well as the numbers of objects in each sample. As expected the
accuracy of photo-z for the $ext$ objects is much better than for
$ptl$. As the sources are extended, they must be at a lower redshift and
the galaxy component non negligible. It is well known that for normal
galaxies the accuracy of photo-z is very high below redshift 1.5 also
with few optical bands that are able to grasp the typical features of
the SED ( i.e. the 4000 \AA ~\  break). On the contrary, point-like
sources are at a higher redshift and the galaxy component is fainter
than the power-law SED describing the bright nucleus. For these
sources, intermediate band photometry would be necessary for
identifying the emission lines and break the degeneracy among the
possible redshift solutions (see Salvato et al 2009, Table 4). For
example, for the $ext$ and $ptl$ samples of objects with 4 or more
bands (4+) we have $\sigma_{\Delta z/(1+z_{sp})}$ = 0.057 and 0.091
with $\eta$ = 11.2 \% and 27.6 \%, respectively. Herewith for the
$all$ sample $\sigma_{\Delta z/(1+z_{sp})}$ = 0.076 with $\eta$ =
22.6 \%.  We did not apply any magnitude cut to our samples but only
6\% of our objects are fainter than $i'$=24.5 and 0.4\% are fainter
than $i'$=26.0 mag. We see that the accuracy of the photometric
redshifts is increasing with the increasing number of bands.

In addition to the computation of the best photometric redshift solution,
LePhare provides the redshift probability distribution (PDZ). For a
large available number of bands  the redshift solution appears as a
clear peak with a PDZ near or equal to 100. When decreasing the number
of bands and an increasing error in photometry, the PDZ presents
either multi-peaks or a unique large distribution of possible
solutions. We used the information in the PDZ for rejecting unreliable
solutions, i.e. when two or more solutions for one object were
produced. From Table 1 we see that if we consider objects with 7 and
more (7+) bands with PDZ=100 then $\sigma_{\Delta z/(1+z_{sp})}$
and $\eta$ are practically the same as for the sample of objects
having 9 or more (9+) bands. However in the case of 9+ bands the number
of objects is twice less than in the case of 7+ bands with
PDZ=100. In the last row of Table 1 we show the accuracy of photo-z
for the magnitude-limited sample with $i'<$22. We see that the accuracy
of photo-z is only negligibly better than for the whole 7+ bands
sample. 

The relations between spectroscopic and photometric redshifts for the
$ext$ and $ptl$ objects which have 7+ bands are presented in the left
and right panels of Fig. 3, respectively. These samples have
$\sigma_{\Delta z/(1+z_{sp})}$=0.057 and 0.087 with a fraction of
outliers  of  10.4\% and 26.7\%, respectively. The number of objects
with PDZ$<100$ does not exceed 41\% of the total number of
objects. Meanwhile the number of outliers among these objects is only
34\%. 

We list hereafter some statistics: 85\% of our $ext$ objects were fitted with
normal galaxy templates (the rest with AGN/QSO hybrid templates) and
74\% of $ptl$ were fitted with templates having an AGN/QSO
contribution (the rest with normal galaxy templates). In Fig. 4 we
show some Seyfert galaxies with clearly visible hosts and bright
AGN. These objects are typical examples of X-ray sources which look
like extended at optical wavelengths and have optical spectra and SEDs
with an AGN/QSO contribution.

Table 2 presents a small fragment of data which is available through
the Milan DB\footnote{http://cosmosdb.iasf-milano.inaf.it/XMM-LSS/}
and the Centre de Donn\'ees de Strasbourg
(CDS\footnote{http://cdsweb.u-strasbg.fr}) for all  5142 considered
objects. 

In this paper we considered the properties of only 3071 objects which
have spectroscopic or photometric redshifts with rank 1 or 2 (see
definitions in the description of columns for Table 2).  Their
redshift distribution is
shown in the bottom panel of Fig. 5. The top panel of Fig. 5 presents
the redshift distribution of 196 GALs and 2875 AGN/QSOs (the final
classification is described in the next subsection). The median
redshifts for the whole sample, GALs and AGN/QSOs are 1.20, 0.19, 1.27,
respectively. We excluded from consideration 52 extended X-ray
objects so they are not represented in Fig. 5.

\begin{figure}
\includegraphics[width=8.5cm,trim=15 15 20 25,clip]{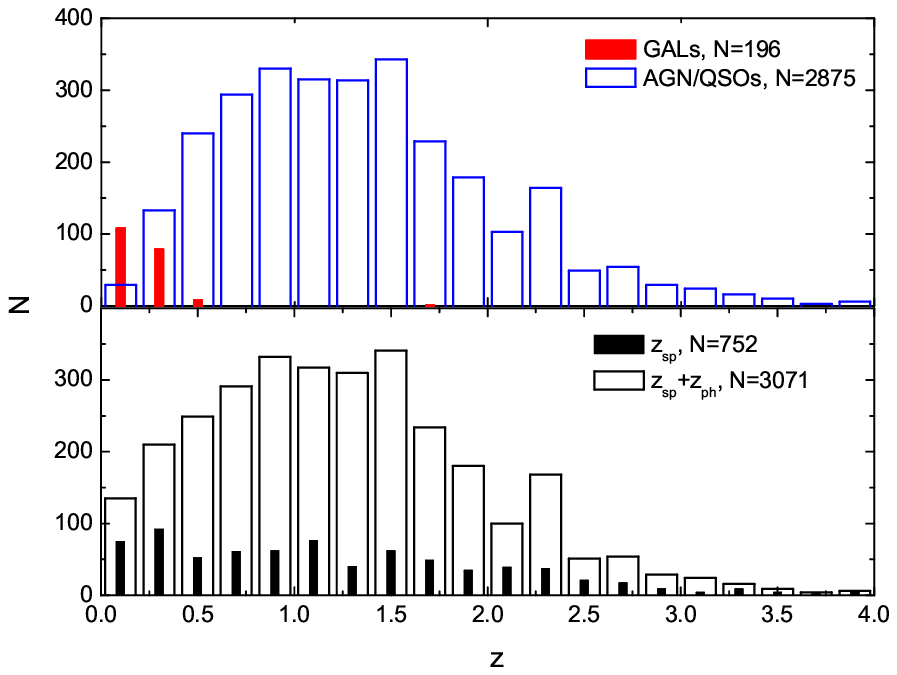}
\caption{{\em Lower Panel:} Redshift distributions of all objects with
  known spectro-$z$ or photo-$z$ (white histogram)  and of objects with spectro-$z$
 (black histogram).
  {\em Upper Panel:} Redshift histogram of sources classified as GALs
  or AGN/QSOs from the z$_{sp}+z_{ph}$ sample.}
\label{5}
\end{figure}

\begin{table*}
\caption{List of an extracted sample of  X-ray sources with optical counterparts
  and their redshifts: just as an example for several sources from the
  main catalog.}
\tabcolsep 2 pt
\begin{tabular}{lccccccccccccccc} \hline

2XLSSd	&	Ora	&	Odec	&	z$_{sp}$	&	zsprank &	zspref	&	zspclass	&	z$_{ph}$	&	sedclass	&	zphrank	&	$HR$	&	$L_{Xsoft}$	&	$L_{Xhard}$	&	final\_class & dogflag	\\
(1)	&	(2)	&	(3)	&	(4)	&	(5)	&	(6)	&	(7)	&	(8)	&	(9)	&	(10)	&	(11)	&	(12)	&	(13)	&	(14) & (15) \\
\hline
J022006.8-045422	&	35.0288	&	-4.9060	&	null	&	null	&	null	&	null	&	0.92	&	gal	&	1	&	-1	&	3.0$\times10^{42}$	&	null	&	AGN/QSO & 0	\\
J022008.7-045906	&	35.0364	&	-4.9849	&	1.65	&	1	&	4	&	AGN	&	1.55	&	agn/qso	&	1	&	-0.64	&	1.3$\times10^{44}$	&	1.9$\times10^{44}$	&	AGN/QSO & 0	\\
J022327.8-040119	&	35.8661	&	-4.0220	&	1.92	&	1	&	2	&	AGN	&	1.88	&	agn/qso	&	1	&	-0.62	&	2.9$\times10^{44}$	&	3.4$\times10^{44}$	&	AGN/QSO & 0	\\
J022500.4-040248	&	36.2523	&	-4.0466	&	0.61	&	2	&	5	&	GAL	&	1.95	&	agn/qso	&	3	&	-1	&	4.5$\times10^{42}$	&	null	&	AGN/QSO & 0	\\
J022624.3-041343	&	36.6016	&	-4.2285	&	null	&	null	&	null	&	null	&	1.57	&	agn/qso	&	2	&	-1	&	1.3$\times10^{44}$	&	null	&	AGN/QSO & 1	\\
J022630.7-040600	&	36.6274	&	-4.0998	&	0.00	&	1	&	4	&	STAR	&	0.00	&	star	&	3	&	-1	&	null	&	null	&	STAR	& 0 \\
J022740.6-041857	&	36.9191	&	-4.3162	&	0.73	&	null	&	6	&	GAL	&	1.74	&	gal	&	2	&	-0.47	&	1.5$\times10^{43}$	&	2.6$\times10^{43}$	&	AGN/QSO & 0	\\
J022758.5-040851	&	36.9936	&	-4.1475	&	1.97	&	2	&	1	&	AGN	&	1.94	&	agn/qso	&	3	&	-1	&	8.2$\times10^{43}$	&	null	&	AGN/QSO & 0	\\
\hline
\end{tabular}

(1) 2XLSSd name; \\
(2) RA of the optical counterpart; \\
(3) DEC of the optical counterpart; \\
(4) spectroscopic redshift when available; \\
(5) rank of the spectroscopic redshift: 1 - good quality (two or more
lines in the spectra), 2 - acceptable redshifts (one clear line in the
spectra) 3 - dubious redshift; \\
(6) source of the spectroscopic redshift: 1 -- Le F\`evre et
al. (2005) , 2 -- Garcet et al. (2007), 3 -- Lacy et al. (2007), 4 --
Stalin et al. (2010), 5 -- Lidman et al. (2012), 6 -- the NASA/IPAC
Extragalactic Database: http://nedwww.ipac.caltech.edu (NED), 7 --
Simpson et al. (2006, 2012), 8 -- Geach et al. (2007), 9 -- Ouchi et
al. (2008), 10 -- Smail et al. (2008), 11 -- van Breukelen et
al. (2007,2009), 12 -- Finoguenov et al. (2010), 13 -- Akiyama et
al. (2013);  14 -- Croom et al. (2013). The compilation of
  redshifts from 7-14 can be found here:
  http://www.nottingham.ac.uk/~ppzoa/UDS\_redshifts\_18Oct2010.fits;
  15 -- SDSS DR9 data: www.sdss3.org. We did not use the SDSS
  redshifts in the calculations; they were added to the table
  later.\\
(7) spectral classification; \\
(8) photometric redshift; \\
(9) classification according to SED: GAL - normal galaxy templates of
Ilbert et al. (2009) or templates \#1-6 of Salvato et. al (2009),
AGN/QSO - hybrid and AGN/QSO templates (\#7-30) of Salvato et al. (2009); \\
(10) rank of the photometric redshift: 1 - good quality photometric
redshift: 7 or more bands, $PDZ=100$; 2 - medium quality photometric
redshift: 7 or more bands, PDZ$<100$; 3 - dubious
photometric redshift, 4-6 bands; 4 - no redshift because of lack of
photometry; 5 - no redshift because the objects are invisible (very
faint) or wrong associations (probably misclassified); \\
(11) hardness ratio $HR$; \\
(12) X-ray luminosity in the soft band, erg/s; \\
(13) X-ray luminosity in the hard band, erg/s; \\
(14) final classification of the source; \\
(15) DOGs objects are flagged by 1 (see subsection 4.3 for the details). 
\end{table*}

\section{X-ray and multiwavelength properties of selected objects}

\subsection{X-ray luminosity}
Fig. 6 shows the X-ray luminosity of the sources, $L_{X}$,  as a function of their
redshift.  The $K$-correction was applied to $L_{X}$ according to
Burlon et al. (2011). 
We mark as "GAL" all spectroscopically confirmed
or photometrically classified galaxies (SEDs fitted by normal galaxy 
templates by Ilbert et al. 2009 or templates  \#1-6 by Salvato et al. 
2009), except for sources with $L_{X, hard} > 2\times 10^{42}$ erg/s  
(or $L_{X, soft} > 10^{42}$ erg/s if the X-ray source was observed 
only in the soft band) which we consider AGN/QSOs as in Brusa
et. al. (2010). We also refer to an object as AGN/QSO if it has a 
spectrum or a SED typical of an AGN or QSO. We have accordingly 
196 (6.4\%) GALs and 2875 (93.6\%)
AGN/QSOs that is in good correspondence with the COSMOS survey 
(6.3\% of X-ray galaxies, Brusa et al. 2010). X-ray luminosities 
and final classification of all considered sources are noted in 
the last column of Table 2.
We have to note that the final sample of GALs consists of 97\% of 
visually classified $ext$ objects and 3\% of $ptl$ while the final 
sample of AGN/QSOs consists of 39\% and 61\% of $ext$ and $ptl$, respectively.

Among the hard X-ray objects (those which have a flux in the 2-10 keV
band) we selected a subsample of sources with
X-ray to optical ratio\footnotemark{} \footnotetext{The X/O ratio was
  calculated as in Brusa et al. (2005) with the $r$-band transformed into
  $R$-band using the empirical Lupton et al. (2005) relation.} X/O$>$10,
as this characteristic indicator is important to identify highly obscured
AGN/QSOs (see for example Fiore et al. 2003, Mignoli et al. 2004, Brusa et
al. 2005, 2010). Our sample consists of 252 highly obscured AGN/QSO
candidates which represents of 11\% of the total AGN/QSO and
 8\% of the whole sample. In Fig. 7 we present the X-ray luminosity vs. $M_{i'}$
absolute luminosity for GALs and AGN/QSOs, where the vertical
solid lines $M_{i'}$=-23.5 divide the AGN from the brighter QSOs. This
limit is very close to the classical definition ($M_{B}$=-23). It is
seen that the Hard ($HR\geq $-0.2) sources (filled symbols) are less
bright than the Soft ($HR<  $-0.2) ones.

\begin{figure}
\includegraphics[width=8.5cm,trim=15 20 30 30,clip]{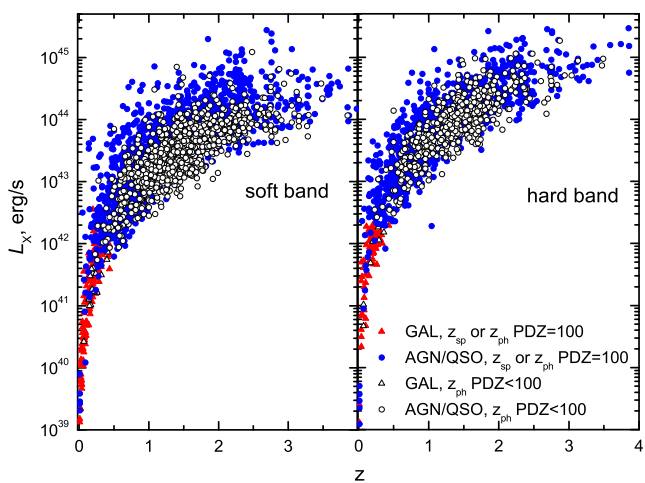}
\caption{X-ray luminosity vs. redshift (photo and spectro) for the
  soft and hard bands.}
\label{6}
\end{figure}

\begin{figure}
\includegraphics[width=8cm,trim=15 10 30 30,clip]{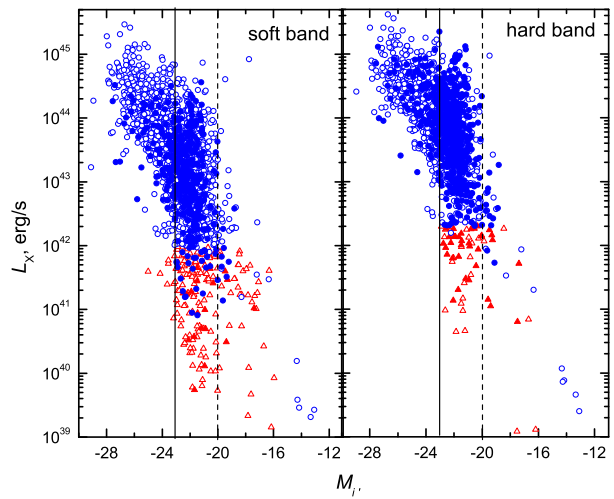}
\caption{ Dependence of X-ray luminosity vs. absolute magnitude in
  the $i'$-band. GALs are marked with triangles, AGN
 with circles, while Soft ($HR<$-0.2) sources are shown with open and Hard ($HR\geq $-0.2)
sources with filled symbols. The solid vertical lines $M_{i'}$=-23.5
divide the AGN from the brighter QSOs. In the overdensity analysis
(section 5) we have only considered the sources locating between the
solid and dashed vertical lines.}
	 \label{7}
      \end{figure}

\subsection{The hardness ratio}

The bottom panel of Fig. 8 presents the comparison of the XMM-LSS
source distributions vs. their hardness ratio $HR=(H-S)/(H+S)$, where
$S$ and $H$ denotes the count rate (cts/s) in the soft and the hard bands,
respectively. Taking into account the average errors of the count
  rates measurements, we estimate a typical uncertainty for $HR$ of
  0.1. We considered distributions of all  the sources from the
X-ray catalog, sources with optical counterparts and sources with
spectro- or photo-z. We see that the various distributions look quite
similar, so we do not see any distinction in the distributions between
X-ray sources showing or not an optical counterpart: the
Kolmogorov-Smirnov probability that distributions are the same is 0.92. 
The middle panel of Fig. 8 presents the distribution of $HR$ for the sources which
were visually classified as extended $ext$ and point-like $ptl$. We may
compare it with the upper panel where the distributions of $HR$ for
GALs and AGN/QSOs are shown. We see that $ext$ sources show some excess of
sources with $HR\geq$-0.2 in comparison with the $ptl$ sample. 

The upper panel of Fig. 8 shows the distributions of $HR$ for GALs and
AGN/QSOs. 
The samples of GAL and AGN/QSOs contain 62\% and 45\%, respectively, 
of the sources observed only in the soft band. The total source sample
contains $\sim$46\% of objects without a hard band detection, a fraction which is 
significantly larger than that of the COSMOS survey ($\sim$ 17\%; Brusa
et al. 2010). We see that some objects which appear as extended at
optical wavelengths have a hard-spectrum and could be considered as
candidates for obscured AGN. We plan to consider the properties of
their host galaxies in a separate work. 

The Kolmogorov-Smirnov probabilities that the $HR$ distributions are
drawn from the same parent population for the pairs of samples $ext$
-- $ptl$  and GAL -- AGN/QSO are less than $10^{-10}$ and $10^{-5}$,
respectively.  The median $HR$ values in the corresponding
  quartile ranges for the All sources, AGN/QSOs and
GALs are -0.63$^{+0.34}_{-0.37}$,  -0.60$^{+0.32}_{-0.40}$ and
-1$^{+0.55}_{-0.00}$, respectively. Note, that 133 GALs out of the
  total of 196 were not observed in the hard band.

It is known that among sources with $HR \geq -0.2 $ there are 80\% of
spectroscopically confirmed obscured QSOs (see for example Ghandi et
al. 2004, Brusa et al. 2010 and
references therein). We have 641 (21\%) of objects with $HR \geq-0.2$
(606 of them are AGN/QSOs) while 154 of them (24\%)  also have X/O
$>$10; these objects are faint at optical wavelengths and very
bright in hard X-rays.

\begin{figure}
\includegraphics[width=8.5cm,trim=15 10 25 15,clip]{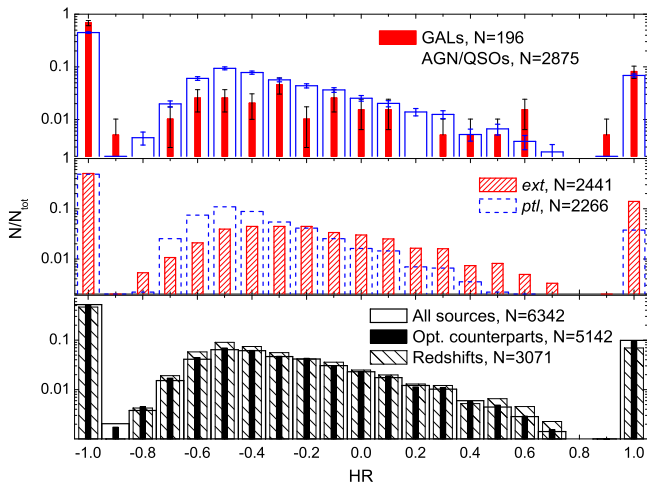}
\caption{Distributions of the fractional number of XMM-LSS X-ray
  sources as a function of their hardness ratio: (i) for the cases of
  All X-ray sources, X-ray sources with optical counterparts, X-ray
  sources with spectro-$z$ or photo-$z$ (bottom panel); (ii) $ext$ and
  $ptl$ objects according to our visual classification
  (middle panel); (iii) GALs and AGN/QSOs according to spectroscopy, SEDs and/or
  L$_{X}$ criteria (top panel).We plot Poisson uncertainties only in
  the upper panel for clarity.}
 \label{8}
\end{figure}

\subsection{Infrared colors}
Lacy et al. (2004) proposed a useful approach for the
classification of AGN with the help of a Spitzer/IRAC color-color
diagram: with AGN/QSOs being concentrated within a compact region of such
a diagram (see also Stern et al. 2005, Lacy et al. 2007, Brusa et al. 2009, 2010). In
total 1467 (51\%) AGN/QSOs and 117 GALs (60\%) were observed in all 4 IRAC
bands. The left panel of Fig. 9 presents the color-color plots for
AGN/QSOs and GALs. We find that 1322 (91\%) of AGN/QSOs and 19 (16\%) of GALs
are lying in the "AGN region".

Dey et al. (2008) and Brodwin et al. (2008) have proposed to select a
population of high-redshift dust-obscured galaxies (DOGs) with large
mid-infrared to ultraviolet luminosity ratios using simple criteria:
$F_{24 \mu m} \geq $ 0.3 mJy  and  $(R-[24]) \geq $ 14 (Vega) mag,
where $[24] \equiv$ -2.5 log$_{10} (F_{24 \mu m}$/7.29 Jy). Using
these criteria, those authors found 2603 DOG candidates in the NOAO
Deep Wide-Field Survey Bootes area, with 86 objects of their sample
having $\langle z\rangle$=1.99 with $\sigma$=0.45.  We have applied
the Dey et al. (2008) criteria and found 54 DOG candidates
which consist of 1.7\% of our sample.  Their average redshift
is lower than that of Dey's objects: $\langle z\rangle$=1.67 with
$\sigma$=0.39. However this value is higher than that
of the total sample $\langle z\rangle$=1.27 with $\sigma$=0.75  (see the $z$
distribution in Fig. 5). In the right panel of Fig. 9 we show the DOGs, 
hard objects with $HR\geq$-0.2 and objects with X/O$>$10. We see
that the DOGs are concentrated in the upper corner of the "AGN
region", while $HR\geq$-0.2 and obscured QSOs with X/O$>$10 are located
in and near the "AGN region". 
Finally we note that out of the 154 $HR\geq-0.2$ and X/O$>10$ objects, 13 are DOGs. 
All DOGs are marked with "1" in the last column of Table 2. 

\begin{figure}
\includegraphics[width=8.5cm,trim=15 15 30 30,clip]{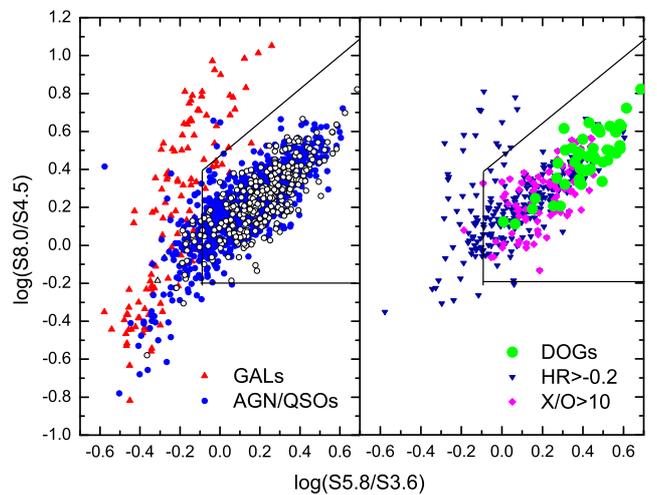}
\caption{IRAC color-color plot. GALs and AGN/QSOs on the left panel (open
  symbols correspond to objects which have photo-$z$ with PDZ$<$100,
  as in Fig. 6). DOGs, HR$\geq$-0.2 and X/O$>$10 sources are shown on the right panel.}
	 \label{9}
      \end{figure}

\section{Environmental properties of the X-ray point-like sources}
We wish to investigate the galaxy environment of our X-ray selected
point-like sources. To this end we will use the CFHTLS W1 optical
object catalog\footnotemark{}
\footnotetext{http://www3.cadc-ccda.hia-iha.nrc-cnrc.gc.ca/community/CFHTLS-SG/docs/cfhtlswide.html\#W1},
but we will only consider "galaxies" with star/galaxy
classification index CLASS\_STAR$<$0.95 and with the $i'$-band limiting
magnitude of 23.5, which is near to the catalog completeness limit \footnotemark{}
\footnotetext{http://terapix.iap.fr/cplt/T0006-doc.pdf}. We have
ignored the sources from the
ABC fields (see Fig. 1) because these areas were not observed in the
$i'$-band. So the considered area in the environmental studies only
concerns $\sim$9 sq. deg.

Taking into account the completeness limits of the CFHTLS and of our
X-ray sample, we investigate the environment of X-ray sources in
a volume limited region of their optical counterparts. This is defined
by the redshift range 0.1$\leq z\leq$0.85 for which their optical
counterparts are within the -23.5$<M_{i'}\leq$-20 absolute magnitude
range (while the rest-frame apparent magnitude of the knee of the $i'$-band
luminosity function has $m^*<$22.5; see Eq. 2). The luminosity
  limits can be seen in Fig. 7 on the $L_X$ vs. $M_{i'}$
  plots.
We consider the following different subsamples of X-ray sources:
{\em All}, GALs (only in the 0.1$\leq z\leq $0.55 redshift range) and
AGN,  Soft ($HR<$-0.2) and Hard ($HR\geq $-0.2) AGN.  We expect that
the overwhelming majority of our Soft AGN are unobscured AGN and the
Hard AGN are obscured ones. 
In Fig. 10 we present the $M_{i'}$ vs. $z$
distribution of {\em All} selected objects (N=777). 
We have to note that the accuracy of the
photometric redshifts for this sample is $\sigma_{\Delta
  z/(1+z_{sp})}$=0.061 with $\eta$=13.8\%. We did not reject from
consideration the objects with PDZ$<$100 because the number of these
objects in our low redshift sample does not exceed 17\% (only 34\% of
objects with PDZ$<100$ are outliers). In any case, we repeated all
calculations, excluding the dubious photo-$z$, and we reached the same
conclusions.

\begin{figure}
\includegraphics[width=8.5cm,trim=15 15 30 30,clip]{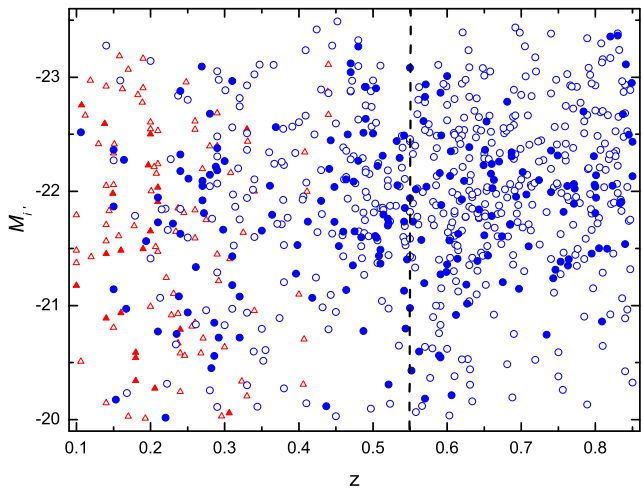}
\caption{Diagram showing the absolute $i'$-band magnitude versus redshift for
  our final volume limited {\em All} sample. GALs are marked with triangles, AGN
 with circles, while Soft ($HR<$-0.2) sources are shown with open and Hard ($HR\geq $-0.2)
sources with filled symbols. The dashed line marks the separation
between two chosen redshift ranges. }
	 \label{10}
      \end{figure}

\subsection{Overdensity measures}
In order to calculate the optical galaxy overdensity around
an X-ray source, we consider concentric annuli centered on each
source (see an example in Fig.11). By taking into account their redshift and angular
distance $D_{A}$, we estimate the linear sizes of the annuli at the
source's rest-frame distance.
Then we count the number, $N_{i}$, of CFHTLS galaxies
within a given annulus in the range of magnitudes from $m^{*}$ to
$m^{*}+\Delta m$ (hereafter {\em fainter} galaxies) or from $m^{*}-\Delta m$ to
$m^{*}$ (hereafter {\em brighter} galaxies),
where $m^{*}$ is the apparent magnitude corresponding to the knee of
the $i'$-band luminosity function [$\Phi(L)$], given by:
\begin{equation}
  m^{*}=5\log_{10} d_{L} + 25 +M_{i'}^{*}+Q_{0.1}(z)+K_{0.1}(z) \;,
\end{equation}
with $M_{i'}^{*} (=-20.82+5\log_{10}h)$ the absolute magnitude of the knee of
the $i$-band $\Phi(L)$ taken from Blanton et al. (2003b),
$Q(z)$ and $K(z)$ the evolution and $K$-corrections, respectively, taken
from Poggianti et al. (1997) and shifted to match their rest-frame
shape at $z=0.1$, $d_{L}$ is the luminosity distance.

\begin{figure}
\includegraphics[width=8.5cm,trim=15 15 30 30,clip]{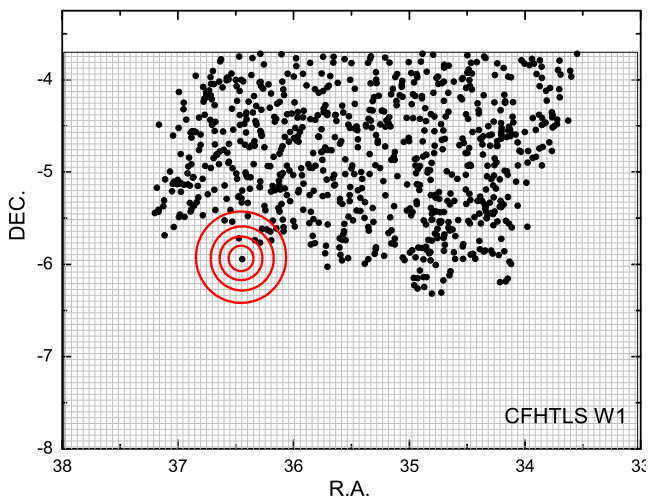}
\caption{The CFHTLS W1 field and corresponding 777 XMM-LSS sources. An
  example of the annuli for one source is shown.}
\label{Fig11}
\end{figure}

Next we have calculated the galaxy overdensities, $\delta_i$, within
each annulus as:
\begin{equation}
\delta_{i}=\frac{N_{i}}{f_{i}N_{b}}-1 \;,
\end{equation}
where $N_{i}$ is the total number of objects within the $i^{th}$ radial
annulus with surface area $A_{i}$, $N_{b}$ is the local background counts,
estimated in the spherical annulus between 3.1 and 5 Mpc, with surface area
$A_{b}$ and $f_{i}$ is the normalization factor that normalizes the background
counts to the area of each spherical annulus. It is given by:
\begin{equation}
f_{i}=\frac{A_{i}}{A_{b}} \;.
\end{equation}
Therefore, for each X-ray source we obtain the overdensity
profile, $\delta_{i}(r)$, as a function of the source-centric distance
$r$. The Poissonian uncertainty of the overdensity, $\delta_{i}$,
is given by:
\begin{equation}
\sigma^{2}_{\delta_{i}}=\frac{1+\delta_{i}}{N_{b}}\left(\frac{1}{f_{i}}+1+\delta_{i}\right)
\end{equation}

In order to have an estimate of the significance of the results, we
compare the overdensity of galaxies around each of our sources with that expected in
mock X-ray source distributions, having random coordinates but
the same fiducial magnitude ($m^{*}$), estimated from the redshift of
the X-ray source itself. For the mock, randomly distributed, sources we
used the same CFHTLS optical catalog as we did for the real ones.

As an example, we find that the mean overdensity of the {\em All} sample within $0.1\leq z \leq
0.55$ and for the first radial annulus is $0.23\pm 0.80$\footnotemark{}
\footnotetext{Here the error is the variance of overdensities
    over the given sample}, 
while for the random distribution the
corresponding value is $0.04\pm 0.75$. Clearly the apparent large scatter 
hinders our ability to distinguish environmental sample differences,
based on the mean overdensity.

As a more sensitive alternative we use a Kolmogorov-Smirnov (KS) two-sample
  test in order to estimate quantitatively the differences between the
  real and random overdensity distributions, constructed for each
  radial distance annulus. 
  In effect we use the cumulative overdensity distribution,
  $$F(>\delta)=N(>\delta)/N_{tot}\;,$$ 
  which is defined as the fraction of all sources ($N_{tot}$)
  having an overdensity above a given $\delta$. For
  the creation of the random overdensity distribution we have generated 100 random catalogs using the procedure described above. 
  For each catalog we calculated the cumulative overdensity distribution $F(>\delta)$. 
  Then we averaged these 100 distributions and obtained the final random distribution which we compared with the real one.
    
  Fig. 12 shows the cumulative overdensity distributions for the real
  and mock samples, for the 0-0.4 Mpc and the 0.4-0.8 Mpc
  radial annuli, for the {\em fainter} ($m^*<m<m^*+1$) and {\em brighter}
  ($m^*-1<m<m^*$) galaxy environments and in the two $0.1\leq z \leq 0.55$ and
  $0.55<z\leq 0.85$ redshift ranges (see labels in each figure). Table
  3 presents the corresponding KS probabilities (${\cal P}$) of the real
  and mock samples being drawn from the same parent population.

\begin{figure*}
\begin{tabular}{cc}
\epsfig{file=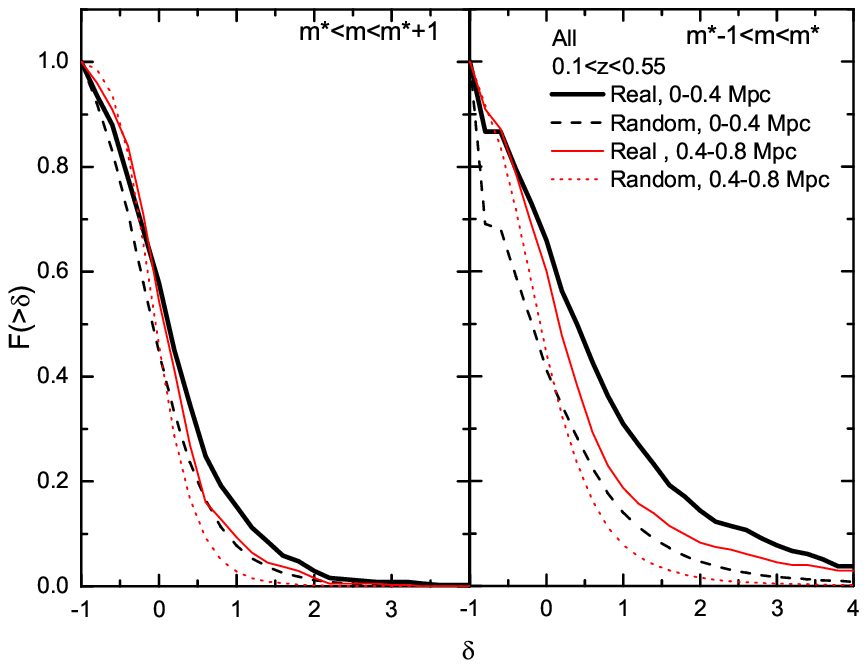,trim=15 15 30 30,clip,width=7.5 cm} &
\epsfig{file=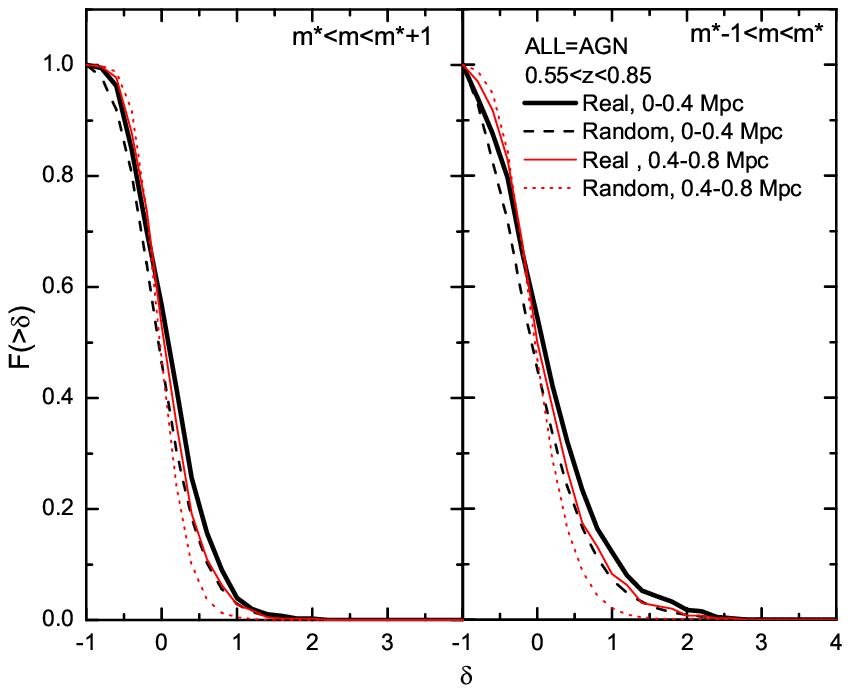,trim=15 15 30 30,clip,width=7 cm} \\
 \epsfig{file=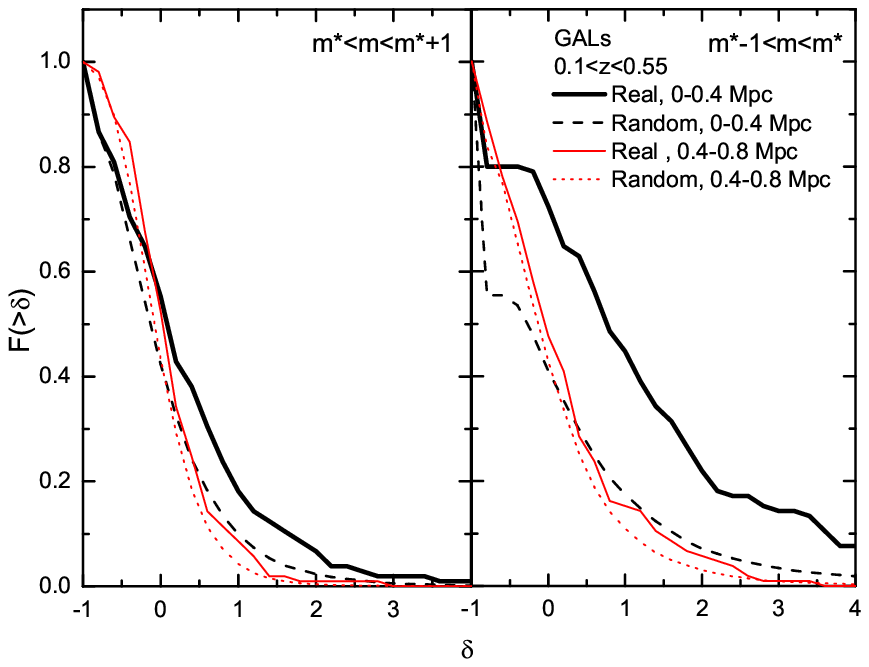,trim=15 15 30 30,clip,width=7.5cm} &
 \epsfig{file=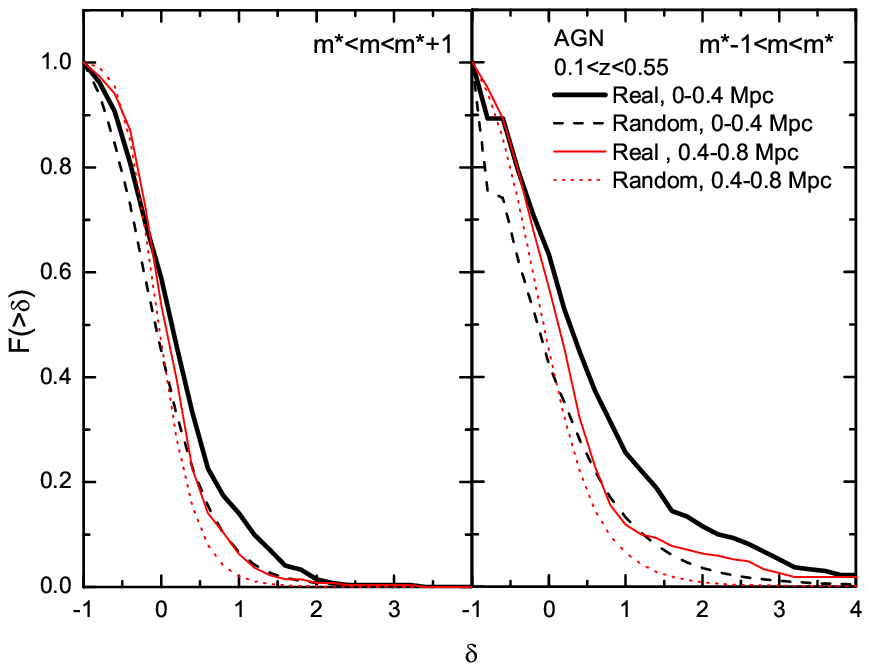,trim=15 15 30 32,clip,width=7.4 cm} \\
 \epsfig{file=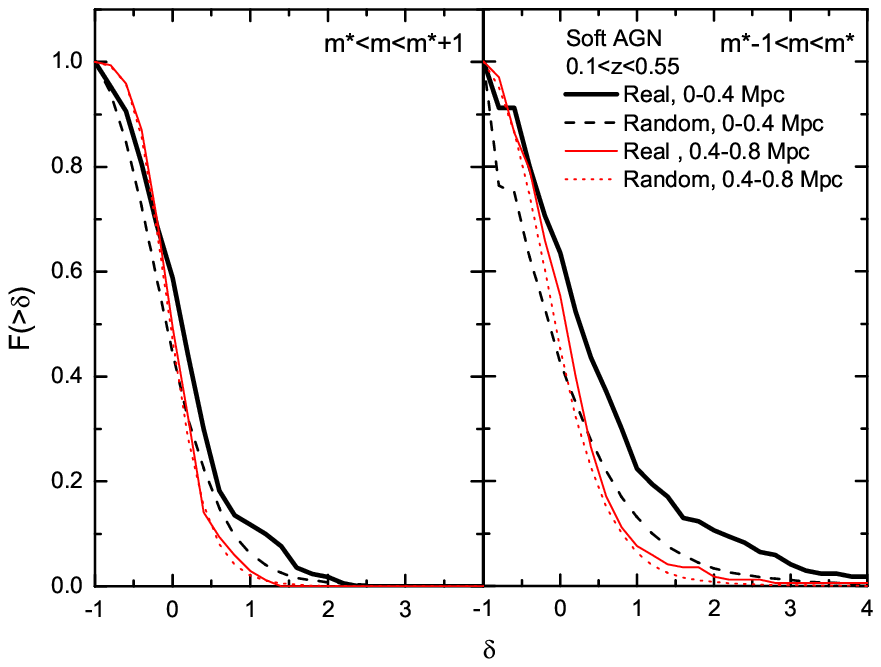,trim=15 15 30 30,clip,width=7.5cm} &
 \epsfig{file=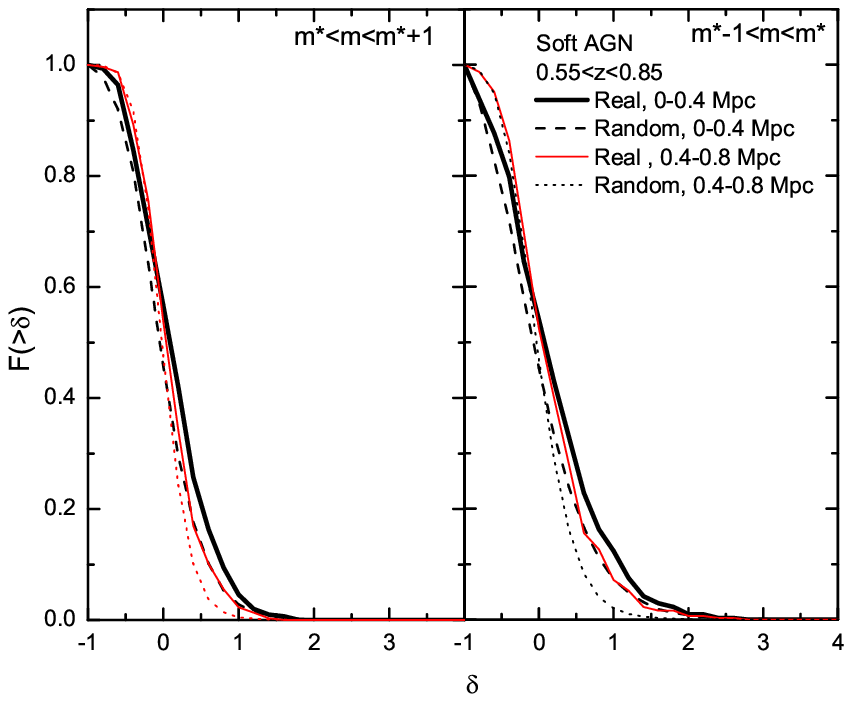,trim=15 17 31 30,clip,width=7.1 cm} \\
 \epsfig{file=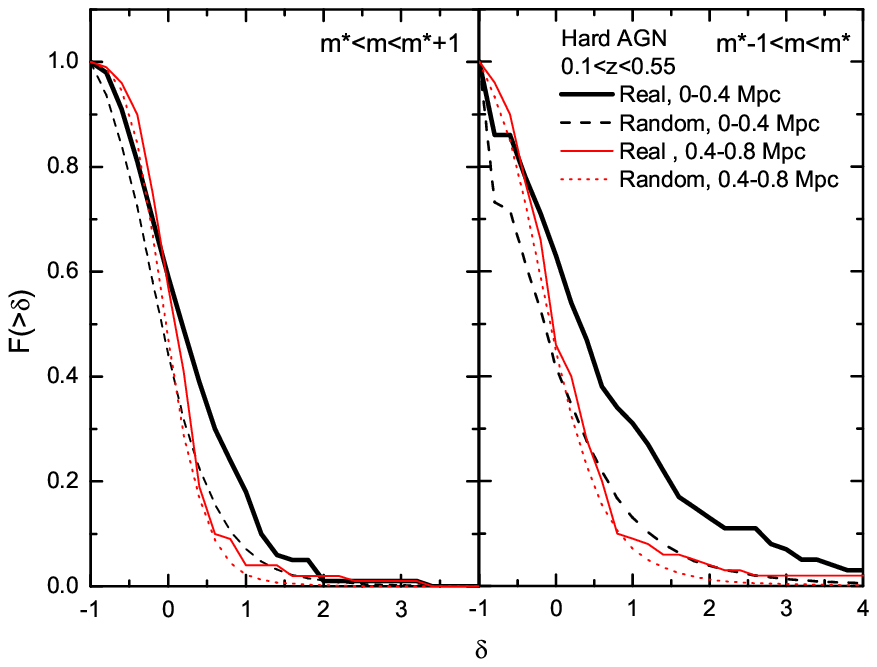,trim=15 15 30 30,clip,width=7.5cm} &
 \epsfig{file=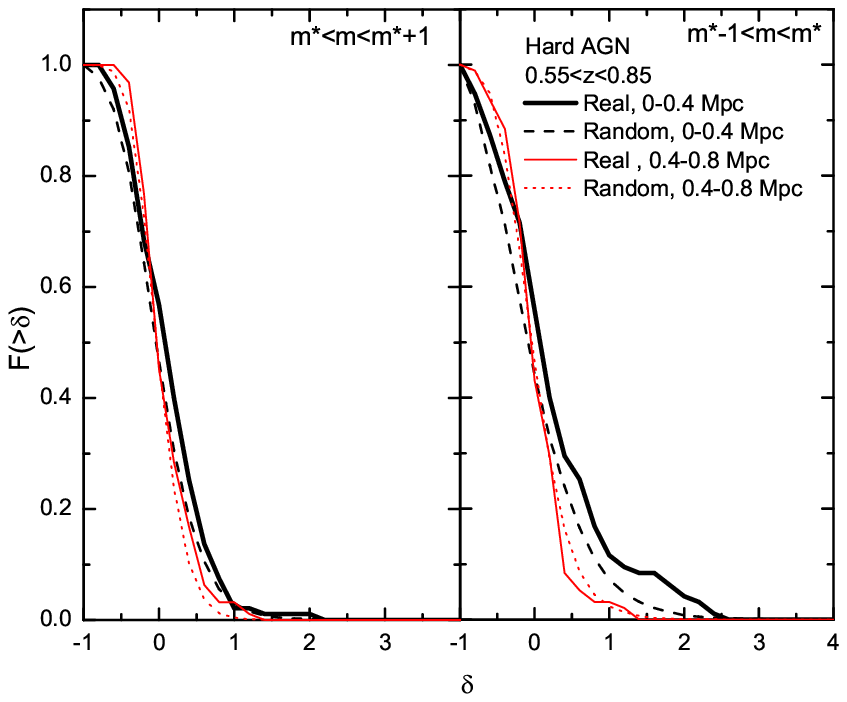,trim=15 15 30 30,clip,width=7 cm} \\
 
 \end{tabular}
\caption{ The cumulative distributions of overdensities for the
    real and mock samples, for the 0-0.4 Mpc and the 
    0.4-0.8 Mpc radial annuli, for the {\em fainter} m*$<$m$<$m*+1 and
    {\em brighter} m*-1$<$m$<$m* environments in the two $0.1\leq z
    \leq 0.55$ (left column) and $0.55<z\leq 0.85$ (right column)
    redshift ranges (see labels in each figure).}
\label{12}
\end{figure*}

\begin{table}
\caption{The Kolmogorov-Smirnov probabilities (${\cal P}$) of the real
  and mock sample overdensity distributions (in the two indicated
  radial annuli) being drawn from the same parent population, for
  the {\em fainter} ($m^*<m<m^*+1$) and {\em brighter} ($m^*-1<m<m^*$) 
  galaxy environments and in the two redshift ranges.}

  \tabcolsep 5 pt

\begin{tabular}{lc|cc|cc} \hline

&  & \multicolumn{2}{c}{$m^*<m<m^*+1$} & \multicolumn{2}{c}{$m^{*}-1<m<m^*$} \\
Sample  & N & ${\cal P}_{0-0.4 Mpc}$ &  ${\cal P}_{0.4-0.8 Mpc}$ & ${\cal P}_{0-0.4 Mpc}$ & ${\cal P}_{0.4-0.8 Mpc}$ \\ \hline \\

\multicolumn{6}{c}{0.1$\leq z \leq$0.55} \\ 

All &375&3.7$\times10^{-6}$&1.3$\times10^{-5}$&1.9$\times10^{-20}$ &1.7$\times10^{-8}$ \\

GAL &105&3.4$\times10^{-2}$&3.5$\times10^{-1}$&1.8$\times10^{-10}$ & 6.1$\times10^{-1}$ \\

AGN &270&4.7$\times10^{-5}$&2.1$\times10^{-3}$ &8.7$\times10^{-11}$&1.6$\times10^{-4}$ \\

Soft AGN&170&1.6$\times10^{-3}$ &9.3$\times10^{-1}$&5.3$\times10^{-7}$ &6.2$\times10^{-2}$ \\

Hard AGN&100&5.0$\times10^{-3}$ &1.0$\times10^{-1}$ &1.9$\times10^{-4}$&6.2$\times10^{-1}$ \\ \\

\multicolumn{6}{c}{0.55$<z\leq$0.85 } \\ 

All&402	&7.6$\times10^{-5}$&6.7$\times10^{-5}$ &1.7$\times10^{-3}$&3.0$\times10^{-4}$\\

Soft AGN&307&1.4$\times10^{-4}$ &4.2$\times10^{-3}$ &6.8$\times10^{-3}$ &3.8$\times10^{-4}$\\

Hard AGN&95&2.6$\times10^{-1}$ &7.8$\times10^{-1}$&4.5$\times10^{-2}$&5.8$\times10^{-1}$ \\ \hline
\end{tabular}

\end{table}

\begin{table}
\caption{The fraction of sources having positive overdensity
    ($\delta>0$) for both the real
  and mock samples (in the two indicated radial annuli); for
  the {\em fainter} ($m^*<m<m^*+1$) and {\em brighter} ($m^*-1<m<m^*$) 
  galaxy environments and in the two studied redshift ranges.}
\tabcolsep 3 pt

\begin{tabular}{l|cc|cc} \hline

&  \multicolumn{4}{c}{$F(\delta>0)\pm\sigma^{*}$, \%} \\
&  \multicolumn{2}{c}{$m^*<m<m^*+1$} & \multicolumn{2}{c}{$m^{*}-1<m<m^*$} \\
Sample   & 0-0.4 Mpc &  0.4-0.8 Mpc & 0-0.4 Mpc & 0.4-0.8 Mpc \\ 
\hline \\

\multicolumn{5}{c}{0.1$\leq z \leq$0.55} \\ 

All  &58$\pm$4&54$\pm$4 &66$\pm$4& 54$\pm$4 \\
All$_{rand}$  &45$\pm$3&46$\pm$4 &41$\pm$3& 44$\pm$3 \\

GAL  &55$\pm$7&52$\pm$7 &72$\pm$8& 52$\pm$7 \\
GAL$_{rand}$  &42$\pm$6&43$\pm$6 &41$\pm$6& 43$\pm$6 \\

AGN  &59$\pm$5&54$\pm$4 &63$\pm$5& 54$\pm$4 \\
AGN$_{rand}$  &45$\pm$4&47$\pm$4 &42$\pm$4& 45$\pm$4 \\

Soft AGN  &59$\pm$6&49$\pm$5 &64$\pm$6& 49$\pm$5 \\
Soft AGN$_{rand}$  &44$\pm$5&47$\pm$5 &43$\pm$5& 45$\pm$5 \\

Hard AGN  &59$\pm$8&57$\pm$8 &63$\pm$8& 57$\pm$8 \\
Hard AGN$_{rand}$  &44$\pm$7&47$\pm$7 &42$\pm$6& 45$\pm$7 \\

\multicolumn{5}{c}{0.55$<z\leq$0.85 } \\ 

All &57$\pm$4&53$\pm$4 &54$\pm$4& 50$\pm$4 \\
All$_{rand}$ &46$\pm$3&47$\pm$3 &45$\pm$3& 47$\pm$3 \\

Soft AGN &57$\pm$4&54$\pm$4 &54$\pm$4& 52$\pm$4 \\
Soft AGN$_{rand}$ &46$\pm$4&48$\pm$4 &45$\pm$4& 47$\pm$4 \\

Hard AGN &57$\pm$8&45$\pm$7 &56$\pm$8& 43$\pm$7 \\ 
Hard AGN$_{rand}$ &47$\pm$7&46$\pm$7 &44$\pm$7& 46$\pm$7 \\
\hline
\end{tabular}
 
 $^{*}$Here the uncertainty represents Poissonian noise.

\end{table}

Although our results show that X-ray point-like sources inhabit, both, dense and
 underdense environments, there are significantly more
 sources inhabiting overdense regions. In all samples we
 find that $F(\delta>0)\magcir 55\%$ for the first radial annuli
 (0-0.4 Mpc), with the random expectation being always $<45\%$ (Table 4).
For example, for the case of the {\em All} sample, in the $0.1\leq z
\leq0.55$ redshift range
 and for the 0-0.4 Mpc radial annulus, we find that the fraction of
 sources with positive overdensity, $F(\delta>0)$, is
58$\pm$4\%/66$\pm$4\% for the {\em fainter/brighter} environments,
while the corresponding random expectation is 45\%/41\%, respectively
(see Table 4 and left top panel in Fig.\ref{12}). 

Furthermore, the KS test show significant differences between
 the {\em All} source overdensity distribution and their random expectations, for both  {\em
   fainter} and {\em brighter} environments, and for the two first
 radial annuli (see Table 3).
In the third radial annulus (0.8-1.2 Mpc), the probability of an
overdensity distribution difference, with respect to the random case,
drops to levels ranging from $\sim 0.05$ to 0.3 for both
type of environments and redshift ranges, and we conclude that at such
large scales we cannot identify significant environmental differences
with respect to the random expectations. Subtle differences could
possibly be revealed only with the use of full redshift information of the
surrounding galaxies.

\begin{figure}
\includegraphics[width=9cm,trim=15 15 20 20,clip]{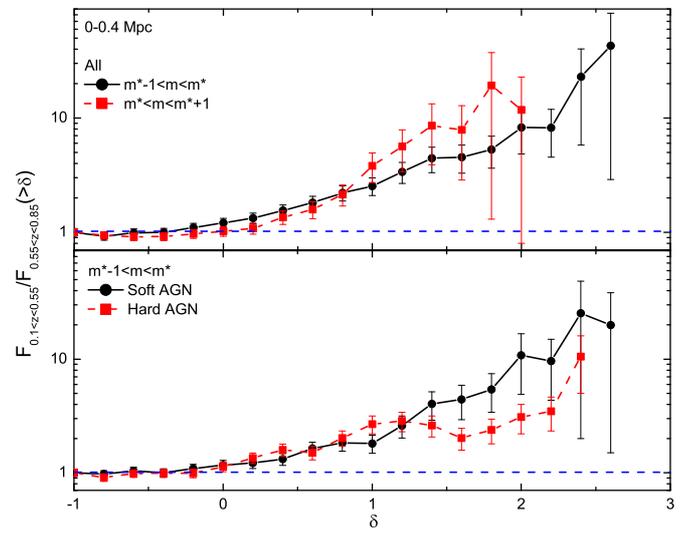} 
\caption{The ratio of the galaxy overdensity distributions between the lower
and higher redshift ranges studied (only in the 0-0.4 Mpc radial
annulus). Errorbars correspond to Poissonian uncertainties.
{\em Upper Panel:} the {\em All} sample. {\em Lower Panel:}
the Hard and Soft AGN samples (only the {\em brighter} environment). }
\label{13}
\end{figure}

An important result of our analysis is that for all our samples 
the overdensities in the 0.1$\leq z \leq$0.55
redshift range are larger and more significant than those in the
$0.55<z\leq 0.85$ range. 
For example, the fraction of {\em All} X-ray sources
having positive overdensities of {\em brighter} galaxies in the 0-0.4 Mpc annulus, 
increases from 54$\pm$4\% to 66$\pm$4\%, between the higher and lower
$z$-ranges.
This effect is clearly seen by inspecting Table 3 and Fig.\ref{13} (upper
panel), where we present the ratio of
the galaxy overdensity distributions between the lower
and higher redshift ranges studied for the {\em All} sample. In the
lower panel of Fig.\ref{13} we present the corresponding ratio separately for the Soft
and Hard AGN sources, which also show a positive evolution of their
galaxy overdensities. A more relevant discussion will be presented further below.

Therefore, there is a positive
redshift evolution of the galaxy overdensity amplitude and/or
significance, within which X-ray sources are embedded.
A similar (weak) tendency was reported by Strand et al. (2008)
for the environments of the optically selected type I quasars.

Again inspecting Table 4 and Fig.\ref{12} we have another, generic, result valid for all
the considered samples, which is that the overdensities defined by the
{\em brighter} galaxies are typically larger and more significant than
those defined by {\em fainter} galaxies. This result should be related to the well known correlation
between galaxy luminosity and clustering amplitude 
(see for example Zehavi et al. 2005, McCracken et al. 2007, Guo et
al. 2013 and references therein).

Furthermore, we can reach a number of
interesting conclusions regarding the environmental differences between
the different source types (AGN, GAL, Soft and Hard AGN).

\begin{figure}
\includegraphics[width=9cm,trim=15 17 20 20,clip]{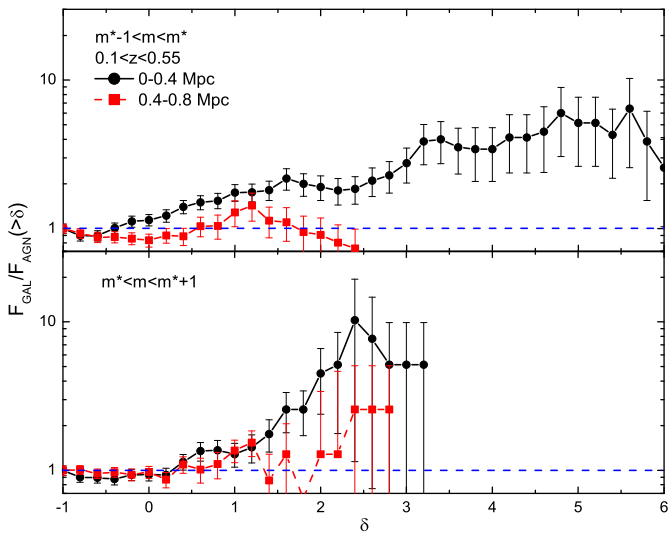}\\ 
\caption{The ratio of the overdensity distributions of the GAL and AGN
samples for both {\em fainter} (lower panel) and {\em brighter} (upper
panel) environments. Black and red curves correspond to the 0-0.4 Mpc
and 0.4-0.8 Mpc annuli, respectively. }
\label{14}
\end{figure}

For example, X-ray galaxies (GAL sample) are found in significantly larger galaxy
overdensities with respect to AGN. This can be clearly seen in Fig.\ref{14} where
we plot the ratio of the overdensity distributions of the GAL and AGN
samples. The excess, by large factors ($\sim 2-10$), of the
positive galaxy overdensities around GAL sources, with respect to those
around AGNs, can be clearly seen but only in the 0-0.4 Mpc annulus
(black line) for both {\em fainter} and {\em brighter}
environments. For the latter type of environment 
we have that $F(\delta>0)$=72$\pm$8\% (with random expectation 41\%) 
and $F(\delta>1)$=45$\pm$7\%, when 
for comparison, the corresponding values for the AGN sample is
$F(\delta>0)$=63$\pm$5\% (with random
expectation 43\%) and $F(\delta>1)$=26$\pm$3\%, respectively. 

\begin{figure}
\includegraphics[width=9cm,trim=15 15 20 20,clip]{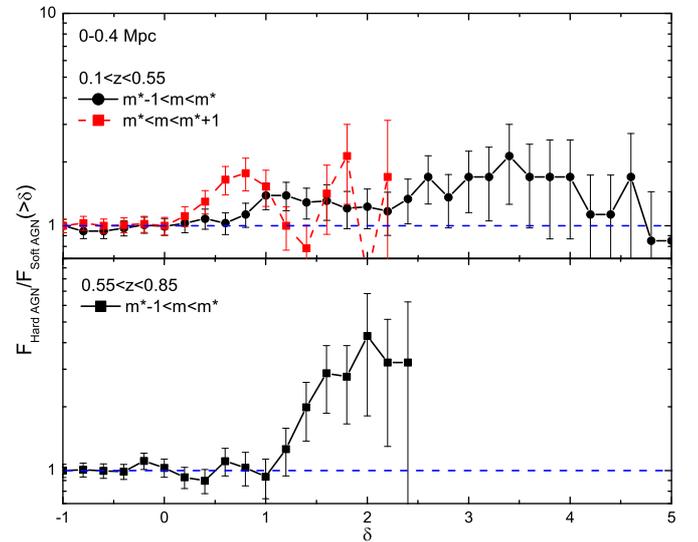} 
\caption{The ratio of the galaxy overdensity distributions of the Hard and
  Soft AGN samples for the 0-0.4 Mpc radial annulus, for 
the {\em fainter} (red dashed line) and {\em brighter} (black
continuous line) environments. Upper/Lower panels correspond to the
lower/higher $z$-range studied. }
\label{15}
\end{figure}

Another interesting, but somewhat unexpected within the unification
paradigm, result is that the Soft and Hard AGN samples
show significant galaxy overdensity distribution differences. 
Inspecting Fig.\ref{15}, where we plot the ratio of the  
overdensity distributions of the Hard and Soft AGN samples, for those
cases where both overdensity distributions are
significantly different than their random expectations,
we see that the Hard AGN sample has always (for $\delta \magcir 0$) a larger
fraction of higher galaxy overdensity values with respect to the Soft AGN's 
(i.e., $F_{\rm Hard}/F_{\rm Soft}(\delta)>1$). This is apparent in both redshift ranges,
although in the higher-$z$ range it appears to be more significant.
This result is in general agreement with the correlation
function analysis of the XMM-LSS sources by Elyiv et al. (2012),
where the clustering of the Hard AGNs was found to be stronger than that of the
Soft AGNs.

Finally, we find another interesting result which is the different
redshift evolution of the galaxy overdensity
distribution for the Soft and Hard sources. In the lower panel of Fig.\ref{13} we plot the
ratio of the overdensity distributions between
the lower and higher redshift ranges, separately for the Soft
(continuous black line) and Hard (red dashed line) AGN in the 0-0.4
Mpc radial annulus. There are indication of a systematic difference by
which  the galaxy overdensities (for $\delta \magcir 1.5$) 
within which the Hard AGN are embedded
evolve less rapidly than the corresponding overdensities
around Soft AGN.

\subsection{Nearest neighbour statistics}
We attempted to investigate the very close environment
around each X-ray source by using a nearest neighbour analysis.
To this end we estimated for each X-ray source the
rest-frame projected linear distance to its nearest optical neighbour
(NNb) using the same CFHTLS galaxy catalog, as in the overdensity
analysis, but within the magnitude ranges $m_{i'}$+$\Delta m$ with
$\Delta m=\pm$ 0.5 or $\pm$1, where $m_{i'}$ is the $i'$-band magnitude of
the host galaxy of the X-ray source. Note, that contrary to the
overdensity analysis for which the characteristic magnitude, $m^*$,
used to define the {\em fainter} and {\em brighter} environment around our X-ray
sources, was that corresponding to the knee of the $i$-band luminosity
function, in the current analysis the characteristic magnitude, $m_{i'}$, is that
of the host galaxy. 

We have then compared, using the KS two-sampled test, the NNb distance
  distribution of our X-ray sources with that of randomly selected CFHTLS
  galaxies having a similar magnitude with that of the X-ray source host
  ($\left| m_{i',X-ray} - m_{i',rand} \right|<0.1$), and found no
  significant difference whatsoever. In Table 5 we present the KS
  probabilities of the indicated pairs of NNb distributions, for
  $\Delta m=+1.0$, -1.0, +0.5, -0.5, being drawn from the same parent
  population. 

It is evident that the only pair of NNb sample distributions that
show a deviation which is marginally significant (${\cal P}_{KS}\simeq
  0.01$) is that of the GAL and AGN 
{\em brighter} neighbors ($\Delta m$=-1.0 and  -0.5). Inspecting the
two distributions (Fig. 16) we see that the 
difference is attributed to a deficiency of GAL neighbours at
a distance of $\sim 0.2$ Mpc and a corresponding excess at
$\sim 0.7$ Mpc, and not to an overall difference in the shape of the
distributions. We consider that the observed difference is not very significant and do not discuss it further.

We have also tested the corresponding
5$^{th}$ nearest neighbour distributions and the results remain the
same.

We conclude that the NNb analysis, applied on our projected data,
cannot be effectively used to characterize differences in the very
close environment of different types of X-ray sources.

\begin{table}
\caption{The Kolmogorov-Smirnov probability of the
  indicated pairs of NNb distributions being drawn from the same
  parent population and for the indicated magnitude range.}
\label{tab1}
\tabcolsep 3pt
\begin{tabular}{ccccc} \hline
 & \multicolumn{4}{c}{${\cal P}_{KS}$}	 \\  
Sample 1/Sample 2  & $\Delta m=+1$ & $\Delta m=-1$  & $\Delta m=0.5$  &$\Delta m=-0.5$ \\ \hline
$0.1\leq z \leq0.55$ & & & & \\
  All/All$_{rand}$ &  0.65 &  0.77 & 0.98 & 0.65 \\
GAL/AGN           & 0.73 & 0.01 & 0.55 & 0.01 \\
Soft AGN/Hard AGN &0.99  & 0.35 & 0.55 & 0.91 \\
\hline
$0.55<z \leq0.85$ & & & &\\
All/All$_{rand}$ &  0.99 & 0.73 & 0.98 & 0.99 \\
Soft AGN/Hard AGN &0.97 &0.99 &0.25 &0.98 \\
\hline
\end{tabular}
\end{table}

\begin{figure}
\includegraphics[width=9cm,trim=15 15 20 20,clip]{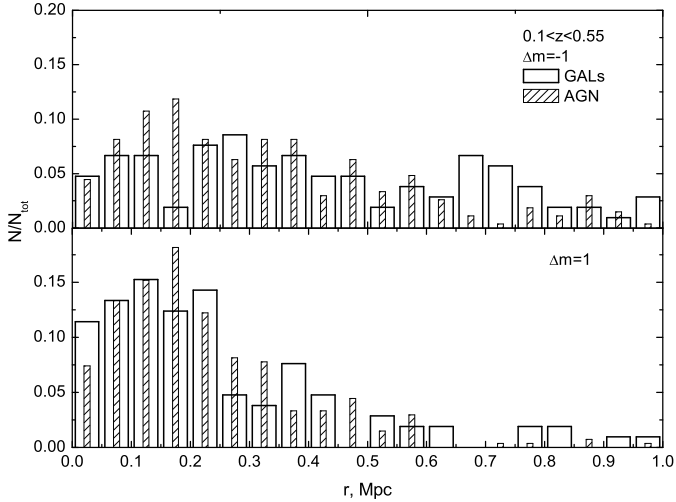} \\
\caption{Differential distributions of the nearest neighbour distance for
  the sample of GALs and AGN for two different magnitude
  ranges in the $0.1\leq z \leq0.55$ redshift range.}
\label{16}
\end{figure}

\section{Summary and main conclusions}
In this paper we have classified 5142 XMM-LSS X-ray
selected sources. In order to check their reliability and positions, reject "problematic"
objects (i.e. defects in the observations, doubtful counterparts, etc.)
and finally classify each object, we visually inspected the optical images of all
X-ray counterparts in the $g'$, $i'$ and $r'$ CFHTLS bands. We classified 2441 objects as being
extended  and 2280 as being point-like, while we rejected from consideration
238 "problematic" objects and 183 stars (5\% and 4\% of the whole sample,
respectively). We estimated the photometric redshifts of the 4435
objects for which there is available photometry in 4-11 bands [i.e., CFHTLS, Spitzer/IRAC,
UKIDSS and GALEX], $i'<$26 mag. Furthermore, 783 objects also have spectroscopic
redshifts. We have found that the photometric redshift accuracy for
the objects with available 4-11 band photometry 
is $\sigma_{\Delta z/(1+z_{sp})}$=0.076 with $\eta$=22.6\%
of outliers. The corresponding values for the objects with 7-11 bands are 
$\sigma_{\Delta z/(1+z_{sp})}$=0.074 with $\eta$=21.8\% and for the
subsample of objects with PDZ=100: $\sigma_{\Delta
  z/(1+z_{sp})}$=0.065 with $\eta$=18.1\%.

We have considered the multiwavelength properties of 3071 objects
which have spectroscopic redshifts or photometric redshifts calculated
from 7-11 bands. 
According to our classification, based on spectra, SEDs and/or
$L_{X}$, we have 196 (6.4\%) GALs and 2875 (93.6\%) AGN/QSOs in our
sample. The median $HR$ values in the corresponding quartile
  ranges for the All sources, AGN/QSOs and
GALs are -0.63$^{+0.34}_{-0.37}$,  -0.60$^{+0.32}_{-0.40}$ and
-1$^{+0.55}_{-0.00}$, respectively. We have also found that 252
objects (8\% of our sample)
have X/O$>$10 and 641 (21\%) have $HR\geq $-0.2, which makes them good
candidates for obscured AGN/QSOs. We have found 54 DOG candidates
(1.7\%) in our sample. 

We have then defined the environment of 777 X-ray
sources (GALs, Soft  and Hard  AGN ($HR<$-0.2 and $HR\geq$-0.2,
respectively) which we considered as unobscured and obscured ones)
with -23.5$< M_{i'}<$-20 in the 0.1$\leq z \leq $0.85 redshift
range. The photo-z accuracy for this low redshift sample is
$\sigma_{\Delta z/(1+z_{sp})}$=0.061 with $\eta$=13.8\%
outliers. Two types of environments have been defined for each X-ray
source; an overdensity of {\em fainter} and of {\em brighter}
galaxies (by one magnitude) with respect to the rest-frame magnitude of the knee of
the $i'$-band luminosity function.
Our main results are the following:

\noindent
(1) {\em The X-ray point-like sources typically reside in overdense
  regions, although they can be found even in underdense regions.} We
find that $\magcir$55-60\% of all X-ray sources
are located in overdense environments ($\delta >$0), a result which is
significantly higher than the random expectation.

\noindent
(2) {\em Overdensities around X-ray sources defined by bright
  neighbours are significantly larger than
  those defined by faint neighbours.} 
For the first redshift
  range $0.1\leq z \leq0.55$ the percentage of objects with positive
  overdensity having  {\em fainter} and {\em brighter} environments
  are 58$\pm$4\% and 66$\pm$4\% against 45\% and 41\% expected for the
  random distribution of the sources, respectively.

\noindent
(3) {\em Overdensities around X-ray sources, defined
  either by brighter or
  fainter neighbours, are typically larger and more significant in
  the $0.1\leq z \leq0.55$ rather than in the $0.55< z \leq0.85$ redshift
  range.} Therefore there appears to be a positive redshift evolution
of the galaxy overdensity amplitude within which X-ray sources are embedded.

\noindent
(4) {\em X-ray galaxies and AGN inhabit different environments.}
X-ray selected galaxies inhabit significantly more overdense {\em brighter} galaxy
regions with respect to AGN, indicating possibly that the former prefer more cluster-like environments,
while the latter group-like ones. However, the overdensities around
X-ray galaxies are significant only up to $\sim 0.4$ Mpc, while those
around AGN up to $\sim 0.8$ Mpc. 
Our results are in general agreement with Georgakakis et al. (2007, 2008),
who showed that X-ray-selected AGN, at $z\sim$1, prefer to reside in
groups and with Bradshaw et al. (2011) who found that
X-ray AGN, in the UDS (SXDS) field with 1.0$<z<$1.5, inhabit significantly
overdense environments corresponding to dark matter haloes of $M\magcir 10^{13} M_{\odot}$.

\noindent
(5)  {\em The obscured AGN ($HR\geq-0.2$) are located in more overdense regions
  with respect to the unobscured AGN ($HR<-0.2$).} This is true for
both {\em brighter} and {\em  fainter} galaxy environments in the 
0.1$\leq z \leq$0.55 redshift range, 
while it is evident only for the {\em brighter} environment in the $0.55<z \leq0.85$
range. This result is in general agreement with the correlation
function analysis of the XMM-LSS X-ray point-source catalog (having
a median $z\sim 1$), presented in Elyiv et al. (2012),
where the clustering of the Hard AGN was found to be stronger than that of the
Soft AGN.
Hickox et al. (2012) also found a
stronger clustering of obscured QSOs with respect to than of
unobscured ones in the 0.7$<z<$1.8 redshift range in the 
Bootes multiwavelength survey, although Allevato et
al. (2011) found an opposite trend.

\noindent
(6) {\em There are some indication that unobscured AGN ($HR<-0.2$) have a
more rapid evolution of their galaxy overdensity amplitude, $\delta$, 
 between the two redshift ranges studied, with respect to the obscured AGN
 ($HR\geq-0.2$).}





\begin{acknowledgements}
We are grateful to Olivier Ilbert for help with LePhare and Mari
Polletta for useful suggestions.

The data used in this work were  obtained
with XMM-Newton, an ESA science mission with instruments and
contributions directly funded by ESA Member States and NASA.

This work is based on observations obtained with MegaPrime/MegaCam, a
joint project of CFHT and CEA/DAPNIA, at the Canada-France-Hawaii
Telescope (CFHT) which is operated by the National Research Council
(NRC) of Canada, the Institut National des Sciences de l'Univers of
the Centre National de la Recherche Scientifique (CNRS) of France, and
the University of Hawaii. This work is based in part on data products
produced at TERAPIX and the Canadian Astronomy Data Centre as part of
the Canada-France-Hawaii Telescope Legacy Survey, a collaborative
project of NRC and CNRS.

We used the data obtained with the Spitzer
Space Telescope, which is operated by the Jet Propulsion Laboratory,
California Institute of Technology under NASA. Support
for this work, part of the Spitzer Space Telescope Legacy Science
Program, was provided by NASA through an award issued by
the Jet Propulsion Laboratory, California Institute of Technology
under NASA contract 1407.

This work is in part based on data collected within the UKIDSS
survey. The UKIDSS project uses the UKIRT Wide Field Camera
funded by the UK Particle Physics and Astronomy Research Council
(PPARC). Financial resources for WFCAM Science Archive
development were provided by the UK Science and Technology
Facilities Council (STFC; formerly by PPARC).

GALEX is a NASA mission managed by the Jet Propulsion
Laboratory. GALEX data used in this paper were obtained from
the Multimission Archive at the Space Telescope Science Institute
(MAST). STScI is operated by the Association of Universities for
Research in Astronomy, Inc., under NASA contract NAS5-26555.
Support for MAST for non-HST data is provided by the NASA
Office of Space Science via grant NNX09AF08G and by other
grants and contracts.

This research has made use of the NASA/IPAC Extragalactic Database
(NED) which is operated by the Jet Propulsion Laboratory, California
Institute of Technology, under contract with the National Aeronautics
and Space Administration.

OM, AE and JS acknowledge support from the ESA PRODEX Programmes "XMM-LSS" and "XXL" and from
the Belgian Federal Science Policy Office. They also acknowledge
support from the Communaut\'e fran\c{c}aise de Belgique - Actions de
recherche concert\'ees - Acad\'emie universitaire Wallonie-Europe.

\end{acknowledgements}

\end{document}